\documentclass[onecolumn]{IEEEtran}

\usepackage{latexsym,url,enumerate}
\usepackage{graphicx}
\usepackage{times,amssymb,amsmath,amsfonts,float,color,bbm,mathrsfs,caption,mathtools,xargs,mathtools,verbatim}
\usepackage{euscript,graphics,cuted}
\usepackage[dvips]{epsfig}
\usepackage[noadjust]{cite}
\usepackage{multirow}
\usepackage{array}
\usepackage[linesnumbered]{algorithm2e}
\usepackage[usenames,dvipsnames]{xcolor}
\usepackage{url}
\usepackage{epsfig}
\usepackage{subfig}
\usepackage{epstopdf}
\usepackage[normalem]{ulem}

\newtheorem{cnstr}{Construction}

\newcommand{\remove}[1]{}

\def\rk{\qopname\relax{no}{rk}}

\newcommand{\ceilenv}[1]{\left\lceil #1 \right\rceil}

\newcommand\nc\newcommand
\nc{\bb}[1]{\mathbb{#1}}
\renewcommand{\cal}[1]{\mathcal{#1}}
\renewcommand{\bf}[1]{\mathbf{#1}}

\DeclarePairedDelimiter{\floor}{\lfloor}{\rfloor}
\DeclarePairedDelimiter{\set}{\lbrace}{\rbrace}
\DeclarePairedDelimiter{\br}{\lparen}{\rparen}

\DeclarePairedDelimiter{\abs}{\lvert}{\rvert}

\nc\bfa{{\bf{a}}}\nc\bfA{{\boldsymbol A}}\nc\cA{{\cal A}} \nc\fA[1]{A\br*{#1}} \nc\fa[1]{a\br*{#1}}  \nc\rmA{\mathrm{A}} \nc\rma{\mathrm{a}}
\nc\bfb{{\bf{b}}}\nc\bfB{{\boldsymbol B}}\nc\cB{{\cal B}} \nc\fB[1]{B\br*{#1}} \nc\fb[1]{b\br*{#1}}  \nc\rmB{\mathrm{B}} \nc\rmb{\mathrm{b}}
\nc\bfc{{\bf{c}}}\nc\bfC{{\boldsymbol C}}\nc\cC{{\cal C}} \nc\fC[1]{C\br*{#1}} \nc\fc[1]{c\br*{#1}}  \nc\rmC{\mathrm{C}} \nc\rmc{\mathrm{c}}
\nc\bfd{{\bf{d}}}\nc\bfD{{\boldsymbol D}}\nc\cD{{\cal D}} \nc\fD[1]{D\br*{#1}} \nc\fd[1]{d\br*{#1}}  \nc\rmD{\mathrm{D}} \nc\rmd{\mathrm{d}}
\nc\bfe{{\bf{e}}}\nc\bfE{{\boldsymbol E}}\nc\cE{{\cal E}} \nc\fE[1]{E\br*{#1}} \nc\fe[1]{e\br*{#1}}  \nc\rmE{\mathrm{E}} \nc\rme{\mathrm{e}}
\nc\bff{{\bf{f}}}\nc\bfF{{\boldsymbol F}}\nc\cF{{\cal F}} \nc\fF[1]{F\br*{#1}} \nc\ff[1]{f\br*{#1}}  \nc\rmF{\mathrm{F}} \nc\rmf{\mathrm{f}}
\nc\bfg{{\bf{g}}}\nc\bfG{{\boldsymbol G}}\nc\cG{{\cal G}} \nc\fG[1]{G\br*{#1}} \nc\fg[1]{g\br*{#1}}  \nc\rmG{\mathrm{G}} \nc\rmg{\mathrm{g}}
\nc\bfh{{\bf{h}}}\nc\bfH{{\boldsymbol H}}\nc\cH{{\cal H}} \nc\fH[1]{H\br*{#1}} \nc\fh[1]{h\br*{#1}}  \nc\rmH{\mathrm{H}} \nc\rmh{\mathrm{h}}
\nc\bfi{{\bf{i}}}\nc\bfI{{\boldsymbol I}}\nc\cI{{\cal I}} \nc\fI[1]{I\br*{#1}} \nc\rmI{\mathrm{I}} \nc\rmi{\mathrm{i}}
\nc\bfj{{\bf{j}}}\nc\bfJ{{\boldsymbol J}}\nc\cJ{{\cal J}} \nc\fJ[1]{J\br*{#1}} \nc\fj[1]{j\br*{#1}} \nc\rmJ{\mathrm{J}} \nc\rmj{\mathrm{j}}
\nc\bfk{{\bf{k}}}\nc\bfK{{\boldsymbol K}}\nc\cK{{\cal K}} \nc\fK[1]{K\br*{#1}} \nc\fk[1]{k\br*{#1}} \nc\rmK{\mathrm{K}} \nc\rmk{\mathrm{k}}
\nc\bfl{{\bf{l}}}\nc\bfL{{\boldsymbol L}}\nc\cL{{\cal L}} \nc\fL[1]{L\br*{#1}} \nc\fl[1]{l\br*{#1}} \nc\rmL{\mathrm{L}} \nc\rml{\mathrm{l}}
\nc\bfm{{\bf{m}}}\nc\bfM{{\boldsymbol M}}\nc\cM{{\cal M}} \nc\fM[1]{M\br*{#1}} \nc\fm[1]{m\br*{#1}} \nc\rmM{\mathrm{M}} \nc\rmm{\mathrm{m}}
\nc\bfn{{\bf{n}}}\nc\bfN{{\boldsymbol N}}\nc\cN{{\cal N}} \nc\fN[1]{N\br*{#1}} \nc\fn[1]{n\br*{#1}} \nc\rmN{\mathrm{N}} \nc\rmn{\mathrm{n}}
\nc\bfo{{\bf{o}}}\nc\bfO{{\boldsymbol O}}\nc\cO{{\cal O}} \nc\fO[1]{O\br*{#1}} \nc\fo[1]{o\br*{#1}} \nc\rmO{\mathrm{O}} \nc\rmo{\mathrm{o}}
\nc\bfp{{\bf{p}}}\nc\bfP{{\boldsymbol P}}\nc\cP{{\cal P}} \nc\fP[1]{P\br*{#1}} \nc\fp[1]{p\br*{#1}} \nc\rmP{\mathrm{P}} \nc\rmp{\mathrm{p}}
\nc\bfq{{\bf{q}}}\nc\bfQ{{\boldsymbol Q}}\nc\cQ{{\cal Q}} \nc\fQ[1]{Q\br*{#1}} \nc\fq[1]{q\br*{#1}} \nc\rmQ{\mathrm{Q}} \nc\rmq{\mathrm{q}}
\nc\bfr{{\bf{r}}}\nc\bfR{{\boldsymbol R}}\nc\cR{{\cal R}} \nc\fR[1]{R\br*{#1}} \nc\fr[1]{r\br*{#1}} \nc\rmR{\mathrm{R}} \nc\rmr{\mathrm{r}}
\nc\bfs{{\bf{s}}}\nc\bfS{{\boldsymbol S}}\nc\cS{{\cal S}} \nc\fS[1]{S\br*{#1}} \nc\fs[1]{s\br*{#1}} \nc\rmS{\mathrm{S}} \nc\rms{\mathrm{s}}
\nc\bft{{\bf{t}}}\nc\bfT{{\boldsymbol T}}\nc\cT{{\cal T}} \nc\fT[1]{T\br*{#1}} \nc\ft[1]{t\br*{#1}} \nc\rmT{\mathrm{T}} \nc\rmt{\mathrm{t}}
\nc\bfu{{\bf{u}}}\nc\bfU{{\boldsymbol U}}\nc\cU{{\cal U}} \nc\fU[1]{U\br*{#1}} \nc\fu[1]{u\br*{#1}} \nc\rmU{\mathrm{U}} \nc\rmu{\mathrm{u}}
\nc\bfv{{\bf{v}}}\nc\bfV{{\boldsymbol V}}\nc\cV{{\cal V}} \nc\fV[1]{V\br*{#1}} \nc\fv[1]{v\br*{#1}} \nc\rmV{\mathrm{V}} \nc\rmv{\mathrm{v}}
\nc\bfw{{\bf{w}}}\nc\bfW{{\boldsymbol W}}\nc\cW{{\cal W}} \nc\fW[1]{W\br*{#1}} \nc\fw[1]{w\br*{#1}} \nc\rmW{\mathrm{W}} \nc\rmw{\mathrm{w}}
\nc\bfx{{\bf{x}}}\nc\bfX{{\boldsymbol X}}\nc\cX{{\cal X}} \nc\fX[1]{X\br*{#1}} \nc\fx[1]{x\br*{#1}} \nc\rmX{\mathrm{X}} \nc\rmx{\mathrm{x}}
\nc\bfy{{\bf{y}}}\nc\bfY{{\boldsymbol Y}}\nc\cY{{\cal Y}} \nc\fY[1]{Y\br*{#1}} \nc\fy[1]{y\br*{#1}} \nc\rmY{\mathrm{Y}} \nc\rmy{\mathrm{y}}
\nc\bfz{{\bf{z}}}\nc\bfZ{{\boldsymbol Z}}\nc\cZ{{\cal Z}} \nc\fZ[1]{Z\br*{#1}} \nc\fz[1]{z\br*{#1}} \nc\rmZ{\mathrm{Z}} \nc\rmz{\mathrm{z}}

\DeclareMathOperator{\Log}{\log}

\DeclareMathOperator{\supp}{supp}

\DeclareMathOperator{\spn}{span}

\DeclareMathOperator{\wt}{wt}

\nc\defeq{\coloneqq}

\DeclarePairedDelimiterX\Set[1]\{\}{#1}

\newcommand\F{{\mathbb F}}

\newtheorem{theorem}{Theorem}
\newtheorem{definition}{Definition}
\newtheorem{lemma}[theorem]{Lemma}
\newtheorem{proposition}[theorem]{Proposition}
\newtheorem{corollary}[theorem]{Corollary}

\newtheorem{remark}{Remark}

\renewcommand{\bfi}{{\underline{i}}}
\renewcommand{\bfj}{{\underline{j}}}

\nc\ellone{{\ell_1}}
\nc\elltwo{{\ell_2}}
\nc\ellinf{{{\ell_\infty}}}
\nc\ip[2]{\langle #1,#2\rangle}
\nc\Clo{\Cl{\rho n}}
\nc\s{\floor*{\frac{r+1}{2}}}
\newcommand{\beeq}{\begin{eqnarray*}}
\newcommand{\eneq}{\end{eqnarray*}}

\newcommand\blfootnote[1]{%
  \begingroup
  \renewcommand\thefootnote{}\footnote{#1}%
  \addtocounter{footnote}{-1}%
  \endgroup
}

\DeclareCaptionLabelSeparator{periodspace}{.\quad}
\newcommand\mmod{\hspace{-.6em}\mod}
\DeclarePairedDelimiterXPP\Cl[1]{B_{\bb{T}}}(){}{\mathopen{}#1}
\newcommandx*\vect[4][1=1, 3=n,4={}]{\begin{pmatrix}#2_{#1}#4\\ \vdots \\ #2_{#3}#4\end{pmatrix}}

\begin{document}
\title{Combinatorial Alphabet-Dependent Bounds for Locally Recoverable Codes}
\author{
Abhishek Agarwal\IEEEauthorrefmark{1}\thanks{\IEEEauthorrefmark{1}College of Information and Computer Sciences,
University of Massachusetts--Amherst, Amherst, MA 01003. Emails: \{abhiag,arya\}@cs.umass.edu. Research supported by NSF grants CCF 1642658, CCF 1318093 and CCF 1618512.},
\and
Alexander Barg\IEEEauthorrefmark{2}\thanks{\IEEEauthorrefmark{2}Dept. of ECE and ISR, University of Maryland, College Park, MD 20742 and IITP, Russian Academy of Sciences,
127051 Moscow, Russia Email: abarg@umd.edu. Research supported in part by NSF grants CCF 1422955 and CCF 1618603.}, 
\and
Sihuang Hu\IEEEauthorrefmark{3}\thanks{\IEEEauthorrefmark{3} Lehrstuhl D f\"ur Mathematik, RWTH Aachen, Germany. Email: husihuang@gmail.com. 
This work was done while this author was a postdoc at Department of Electrical Engineering - Systems, Tel Aviv University, Israel. 
Research supported by ERC grant no.~639573, ISF grant no.~1367/14, and the Alexander von Humboldt Foundation.},
%\hspace{0.7cm} 
\and 
Arya Mazumdar\IEEEauthorrefmark{1},
%\thanks{\IEEEauthorrefmark{4}College of Information and Computer Sciences,
%University of Massachusetts Amherst, Amherst, MA 01003. Email: arya@cs.umass.edu. Research }
\and
Itzhak Tamo\IEEEauthorrefmark{4}
\thanks{\IEEEauthorrefmark{4} Department of Electrical Engineering - Systems, Tel Aviv University, Israel. Email: tamo@post.tau.ac.il. Research supported by ISF grant no.~1030/15 and NSF-BSF grant no.~2015814.}
}

\maketitle

\begin{abstract}
Locally recoverable (LRC) codes have recently been a focus point of research in coding theory due to their theoretical appeal and applications in distributed storage systems. In an LRC code, any erased symbol of a codeword can be recovered by accessing only 
a small number of other symbols. For LRC codes over a small alphabet (such as binary), the optimal rate-distance trade-off is unknown. 
We present several new combinatorial bounds on LRC codes including the locality-aware sphere packing and Plotkin bounds. We also develop an approach to linear programming (LP) bounds on LRC codes. The resulting LP bound gives better estimates in examples than the other upper bounds known in the literature.
Further, we provide the tightest known upper bound on the rate of linear LRC codes with a given relative distance, an improvement over the previous best known bounds. % by Cadambe and Mazumdar.
%A locally recoverable (LRC) code is a code that enables simple recovery of an erased symbol by accessing only a small number of other symbols. 
\end{abstract}

\section{Introduction}
\blfootnote{The authors' names appear in alphabetical order.}
\blfootnote{This paper was presented in part at 2016 IEEE International Symposium on Information Theory, Barcelona, Spain, July 2016, and at 54th Annual Allerton Conference on Communication, Control and Computing, Monticello, IL, September 2016.}
 
We consider codes over a finite alphabet that have the usual property of error correction and the additional property of
being able to recover one or more erased symbols of the codeword by accessing only a small number of other symbols. Codes
of this kind are said to be locally recoverable (LRC), and they have applications in large-scale distributed storage systems. 
LRC codes were first defined in \cite{Gop11} and were studied in a number of subsequent papers in recent years.

A $q$-ary code $\cC$ of length $n,$ cardinality $M,$ and  distance $d$ is a set of $M$ vectors over an alphabet $Q, |Q|=q$ with minimum pairwise Hamming distance $d$. The quantity $k=\log_qM$ is called the dimension of $\cC.$ If $Q$ is a finite field and
$\cC$ is a linear subspace of $Q^n$ then $k$ is the dimension of $\cC$ as a vector space. Below, $[n] \equiv \{1,\dots, n\}$, and for any $x \in Q^n$, $x_{i}$ is the projection of $x$ in the $i$th coordinate. By extension, for any $I \subseteq [n]$, $x_I$ is the projection of $x$ onto the coordinates of $I$.

\begin{definition}\label{def1}
%A code $\cC$  of length $n$ is said to have {\em locality $r$} if every coordinate $i\in [n]$ is contained in a 
%subset $\cR_i\subset[n]$ of size at most $r+1$ with the property that there exists a function $\phi_i$ 
%such that for every codeword $c\in \cC$,  
%   \begin{equation}\label{eq:Ri1}
%     c_i=\phi_i(\{c_j, j\in \cR_i\backslash \{i\}\}).
%   \end{equation}
  A code $\cC\subset Q^n$ is \emph{locally recoverable with locality $r$}
if every coordinate $i\in \{1,2,\dots,n\}$ is contained in a subset $\cR_i\subseteq[n]$ of size $r+1$ such that
for every codeword $c\in\cC$ there is a function
 $\phi_i:Q^r\to Q$ with the property that
   \begin{equation}\label{eq:def1}
   c_i=\phi_i(c_{j_1},\dots,c_{j_r}),
   \end{equation}
where $j_1 < j_2 < \cdots < j_r$ are the elements of $\cR_i\backslash\{i\}.$
We use the notation $(n,k,r)$ to refer to a code of length $n$, dimension $k$ and locality $r.$
\end{definition}
The definition of LRC codes was extended in several different ways. The following generalization  is important for
our purposes.
\vspace*{.05in}
\begin{definition} A code $\cC\subset Q^n$ of cardinality $q^k$ is said to have the $(\rho,r)$ {\em locality property} (to be an $(n,k,r,\rho)$ LRC code) where $\rho\geq 2$, 
if each coordinate $i\in [n]$ is contained in a subset $\cR_i\subset [n]$ of size at most $r+\rho -1$ 
such that the restriction $\cC_{\cR_i}$ of the code $\cC$ to the coordinates in $\cR_i$ forms a code of distance at least $\rho$. 
Notice that the values of any $\rho-1$ coordinates of $\cR_i$ are determined by  the values  of the  remaining $|\cR_i|-(\rho-1)\leq r$ coordinates, thus enabling local recovery. 
\label{def:LRC}
\end{definition}
\vspace*{.05in}
This definition was first proposed in \cite{PKLK2012,kamath2013} with the less demanding restriction of protecting only the information 
symbols of the codeword (see also \cite{Lluis2013} for a related but different notion). In the above definition we 
consider all-symbol locality, without differentiating between the information and parity symbols. The set $\cR_i$ is called
the {\em repair group}, and the set $\cR_i\backslash\{i\}$ is called the {\em recovery set} for the coordinate $i$.

Other extensions of the concept of LRC codes include codes with multiple disjoint repair groups for every coordinate, also 
called codes with the availability property \cite{wang2014a}, codes with sequential repair of several erasures \cite{Prakash14},
codes with cooperative repair \cite{RawatEurasip15}, local repair on graphs \cite{mazumdar2015storage}, as well as other variations. 

Problems of constructing LRC codes and bounding their parameters have been the subject of a 
considerable number of publications. Constructions of LRC codes obtained by combining some known code families without the
locality property were suggested in \cite{sil13,huang2007pyramid,TPDMatroids}. A family of codes extending the construction of
 Reed-Solomon codes to codes with locality was proposed in \cite{Tamo13} and further generalized to codes on algebraic curves in \cite{BTV17}.
We refer to \cite{barg15} for a survey of some aspects of the algebraic theory of LRC codes.

Research on bounds for LRC codes was
initiated in \cite{Gop11} which 
  showed that the distance $d(\cC)$ of an $(n,k,r)$ LRC code $\cC$ is bounded as follows:
  \begin{equation}\label{eq:msft}
  d(\cC)\le n-k-\Big\lceil\frac kr\Big\rceil +2.
  \end{equation}
In \cite{kamath14} this bound was extended to the case of arbitrary $\rho\ge 2$. Namely, the distance of an $(n,k,r,\rho)$
LRC code satisfies the inequality
  \begin{equation}\label{eq:rho}
  d\le n-k+1-\Big(\Big\lceil\frac kr\Big\rceil-1\Big)(\rho-1).
  \end{equation}
Bounds for codes with availability were established in \cite{wang2014a,raw14,TBF16}.

Note that the bounds \eqref{eq:msft},~\eqref{eq:rho} do not depend on the size of the code alphabet $q$. A bound that 
accounts for the value of $q$ was derived in \cite{CaMa2015}. It has the following form: For any 
$q$-ary LRC code with the parameters $(n,k,r)$ and distance $d,$
  \begin{equation}\label{eq:cm}
    k\le \min_{1\le s\le n/(r+1)} \,\{sr+\log_q M_q(n-s(r+1),d)\},
  \end{equation}
where $M_q(n,d)$ is the maximum cardinality of a $q$-ary length $n$ code with distance $d$.
This bound can be used to derive asymptotic upper bounds on the rate of LRC codes with a given value of the distance (more on this
below).
Asymptotic lower bounds (achievability results) on the rate of LRC codes, namely Gilbert-Varshamov (GV) type asymptotic bounds, were 
also derived independently in \cite{TBF16,CaMa2015}; in particular the former work derives a  bound for the case of availability $2$ as well. 

In this paper we focus on combinatorial upper bounds on the parameters of LRC codes, tightening prior results, and emphasizing the 
dependence between the parameters and the size of the code alphabet $q$. We explore several general approaches to the derivation
of the upper bounds, including recursive bounds, the linear programming approach, and the approach relying on the coset leader graph of the 
code. 
%A recursive bound \eqref{eq:cm} is derived using a shortening argument. We find a different kind of recursion that enables us to use known bounds on codes with no locality assumption to derive bounds on LRC codes. 

Linear programming (LP) is a powerful technique that accounts for some of the best known upper bounds on the size of codes with a 
given distance. It was pioneered in \cite{del73} and used in \cite{mce77} to derive the best currently known
asymptotic upper bound on error correcting codes. These results rely on the approach to codes via association schemes and 
their eigenvalues, combined with some analytic techniques. Incorporating the locality constraints into the LP problem in a way that yields closed-form bounds is a nontrivial problem. We suggest a way to address it under the additional assumption that 
$\cR_i\cap \cR_j = \emptyset, i\ne j,$ i.e., that different repair groups are disjoint, and the set of coordinates $[n]$ is a disjoint
union of the repair groups. With this assumption,
an association scheme that fits the locality constraints forms a Delsarte extension of the usual Hamming scheme.
Relying on this observation, we derive an LP bound on $(n,k,\rho,r)$ LRC codes
in a polynomial form and construct a polynomial that gives rise to a Singleton-like bound on such codes. We also compute numerical examples for $\rho=2,$
which corresponds to the original definition of LRC codes, and show that the LP bound is sometimes better than the only other known alphabet-dependent bound \eqref{eq:cm}.

We note that LP bounds on linear LRC codes were earlier studied in \cite{ChanLP13} which considered a standard LP problem \cite{del73} with the additional constraint that every coordinate is contained in a codeword of the dual code of weight $\le r+1.$ At the same time \cite{ChanLP13} gave no closed-form solutions of the LP problem or any numerical examples. 
LP bounds for LRC codes with multiple repair groups were considered in~\cite{TebbiChanSung2014} and LP bounds for cyclic LRC codes were considered in \cite{CyclicLRC16}. 

Finally, we study asymptotic upper bounds on linear LRC codes that satisfy Definition
\ref{def1}. The starting point of our study is an observation that a linear LRC code necessarily contains several low-weight 
parity checks. Another class of codes that has the same property is low-density parity check (LDPC) codes. A recent work
\cite{iceland2015coset} derived new improved asymptotic bounds on the rate of LDPC codes by analyzing the coset graph of the code.
While LDPC codes by definition contain only low-weight parity checks, LRC codes combine such checks with a large number of 
unrestricted parity check equations. Nevertheless, it is possible to combine the approach of \cite{iceland2015coset} with the
recursive bound \eqref{eq:cm} to obtain
an asymptotic bound on linear LRC codes that is better than the asymptotic bound obtained from \eqref{eq:cm}.
An even better bound can be obtained for linear LRC codes with disjoint repair groups.

The paper is organized as follows.
In Section \ref{sect:cb} we derive a general upper bound on the size of LRC codes that reduces the problem to bounds on codes with a given distance but without locality constraints. This result is conceptually similar to the bound \eqref{eq:cm} (from \cite{CaMa2015}) but relies on a different kind of recursion, and this reduction enables us to use known bounds on codes to derive new results for LRC codes. In conjunction with the asymptotic Gilbert-Varshamov (GV) bound on $(n,k,r,\rho)$ LRC codes of \cite{BTV17}, this yield an exact value of the asymptotic code rate when $d/n\to 0.$ This result, proved earlier for $\rho=2$ in \cite{CaMa2015,TBF16}, is extended here to any $\rho\ge 2.$
In Section \ref{sect:LP}, we derive Delsarte's linear programming (LP) bounds for LRC codes with disjoint repair groups. The results include a Singleton bound for LRC codes. For the special case of usual LRC codes (Definition \ref{def1}) with disjoint repair groups, our results improve the shortening bound of  \eqref{eq:cm}.
In Section~\ref{sec:asym}, we consider linear LRC codes, and by using a theory of coset-leader graphs combined with the approach of \cite{CaMa2015},
are able to provide better asymptotic bounds on the rate-relative distance trade-off of LRC codes. The bound becomes stronger if we consider disjoint repair groups.

\textcolor{black}{This paper is a result of merging and developing the papers by S. Hu, I. Tamo, and A. Barg  \cite{HTB16} and  by
A. Agarwal and A. Mazumdar \cite{AM16}, both devoted to the problem
of deriving alphabet-dependent bounds on LRC codes.}

\section{New Bounds on LRC Codes}\label{sect:cb}
In the next theorem we introduce a method of using upper bounds on codes with a given distance (without the locality property) to derive upper bounds on LRC codes.
{Let $B(l,\rho)$ be an upper bound on the cardinality of a code of length $l$ and distance $\rho$, which is a 
log-convex function\footnote{A positive function $f(j)$ of the integer argument is called log-convex if 
$f(j_1)f(j_2)\le f(j_1-1)f(j_2+1)$ for 
any $j_1\leq j_2$ in the support of $f$.}
of $l$ and such that $B(0,\rho)=1.$ 
}

\begin{theorem}\label{thm:RecursiveUpperBound}
%\sout{Let $B(l,\rho)$ be an upper bound on the cardinality of a code of length $l$ and distance $\rho$, which is a 
%log-convex function of $l$ and such that $B(0,\rho)=1.$}
Let $\cC$ be an $(n,k,r,\rho)$ $q$-ary LRC code with distance $d$, and let 
  \begin{equation}\label{eq:mu}
  {\color{black}\mu=\mu(n,d,r,\rho):=\ceilenv{ \frac{n-(d-1)}{{N}}}+1},
  \end{equation}
where $N= r+\rho-1$.
Then, for any $\rho\ge 2$, we have
   \begin{equation}\label{main-thm}
   k\leq {\mu}\log_q B({N},\rho).
   \end{equation}
\end{theorem} 

%\begin{remark} Suppose that the code $\cC$ is a direct sum of codes of length ${N},$ then the bound \eqref{main-thm} is obvious.
%The proof below sidesteps the structural condition at the expense of an additional assumption.
%\end{remark}

\begin{IEEEproof}
We begin with constructing a sequence of nonempty disjoint subsets $X_i\subset [n]$ whose union is of size at least $n-(d-1).$ 
Starting with $X_0=\emptyset,$ assume that the sets $X_0,...,X_{i},i\ge 0$ are already constructed. 
If $|\cup_{l=0}^{i}X_l|\geq n-(d-1),$ terminate the procedure. 
Otherwise let $j$ be an arbitrary element in $[n]\backslash \cup_{l=0}^i X_l$ 
(w.l.o.g. we can assume that $j=i+1$) and define
%$$X_{i+1}=\cR_j\backslash \cup_{l=0}^i X_l.$$ 
%W.l.o.g. we can assume that $j=i+1$ and 
$$X_{i+1}=\cR_{i+1}\backslash \cup_{l=0}^i X_l.$$
Suppose that this procedure terminates after $m$ steps, and let $X_1,...,X_m$ be the sequence of subsets constructed above.  
For $i=1,...,m$ let $X_{[i]}:=\cup_{l=1}^iX_l$ and denote by  $\cC_i$  the restriction of $\cC$ to the coordinates in $X_{[i]}$. Note that by the construction, we have 
$$ 
|X_{[m-1]}| < n-(d-1),
$$
and 
\begin{equation} \label{eq:stam0}
  |X_{[m]}| \le |X_{[m-1]}|+{N} \le \mu{N}.
\end{equation}  

Let us prove by induction on $i$ that 
  \begin{equation}\label{eq:i}
|\cC_i|\le \prod_{j=1}^iB(|X_j|,\rho) \text{ for all }i=1,...,m.
  \end{equation}
   For $i=1$, $X_1=\cR_1$ 
and by  definition,  the code $\cC_{1}=\cC_{\cR_1}$ has distance at least $\rho.$ Therefore,  
$|\cC_1|\le B(|X_1|,\rho)$.   
Now assume that \eqref{eq:i} holds for $\cC_{i-1}.$ Let $c$ be an arbitrary codeword of $\cC_{i-1}$, and let $S(c)$ be the set of codewords in $\cC_i$ whose restriction to $X_{[i-1]}$ equals $c$. These codewords can be different only in the coordinates
in $X_i\ (\subseteq \cR_i)$, and therefore the restriction of $S(c)$ to the coordinates in $X_i$ forms a code of distance at least $\rho.$ This implies that $|S(c)|\le B(|X_i|,\rho)$, and so $|\cC_i|\le |\cC_{i-1}|B(|X_i|,\rho).$
This completes the induction step.

Since the code $\cC$ has distance $d$ and $|X_{[m]}|\ge n-(d-1)$, it follows that 
\begin{equation}
|\cC|=|\cC_m|\leq \prod_{j=1}^mB(|X_j|,\rho).
\label{eq:stam}
\end{equation}
Suppose that $i,j$ are such that  $1\le |X_i|\leq |X_j|\le {N},$ then using log-convexity, we obtain 
   \begin{align*}
   B(|X_i|,\rho) B(|X_j|,\rho)&\leq B(|X_i|-1,\rho)  B(|X_j|+1,\rho).
   \end{align*}
This step can be repeated $\min(|X_i|, {N}-|X_j|)$ times till either the larger subset is of the maximum possible size ${N}$ or
the smaller one becomes empty (in which case we put $B(0,\rho)=1$).
Use this argument in \eqref{eq:stam} and successively reduce the number of factors on the right-hand side as many times as possible.
On account of \eqref{eq:stam0} we will obtain at most $\mu$ factors, and in each of them the size of the coordinate subset $X_{\{\cdot\}}$
will be ${N}$ or less.
We conclude that
\begin{equation}
\prod_{j=1}^mB(|X_j|,\rho)\leq B({N},\rho)^{{\mu}}.   
\label{eq:stam2}
\end{equation}
Now \eqref{main-thm} follows by combining \eqref{eq:stam} and \eqref{eq:stam2} and taking logarithms on both sides of the resulting equation.
\end{IEEEproof}

\vspace*{.1in}
Theorem~\ref{thm:RecursiveUpperBound} provides a general upper bound on the size of LRC codes. Explicit results are obtained once we
substitute a log-convex upper bound $B(\cdot)$. Fortunately, many known bounds on codes are in fact log-convex. For instance, let us prove that this is the case for the Hamming (sphere-packing) bound.
\begin{lemma}
The function $B_H(n,e)={q^n}/{(\sum_{i=0}^e \binom ni (q-1)^i)}$ is log-convex in $n$.
\end{lemma}
\begin{IEEEproof} Let ${Q}=q-1$ and let $f(n,e)=\sum_{i=0}^e{n\choose i}{Q}^i$. In all the expressions below $e$ does not change, so 
to simplify the notation we write $f(n)$ instead of $f(n,e)$. For any $n\ge 1$ we have
   \begin{align*}
   f(n)&=\sum_{i=0}^e \Big[\binom{n-1}i+\binom{n-1}{i-1}\Big]{Q}^i\\
         &=(1+{Q})f(n-1)-\binom{n-1}e{Q}^{e+1}.
   \end{align*}
We find
  \begin{align*}
  f(n_1)&f(n_2)-f(n_1-1)f(n_2+1)\\
  =&\Big[(1+{Q})f(n_1-1)-\binom{n_1-1}{e}{Q}^{e+1}\Big]f(n_2)\\
   &-f(n_1-1)\Big[(1+{Q})f(n_2)-\binom{n_2}e{Q}^{e+1}\Big]\\
   %&\hspace*{0.5in}  -f(n_1-1)\Big((1+{Q})f(n_2)-\binom{n_2}e{Q}^{e+1}\Big)\\
  =&\Big[\binom{n_2}{e}f(n_1-1)-\binom{n_1-1}{e}f(n_2)\Big]{Q}^{e+1}\\
  =&{Q}^{e+1}\sum_{i=0}^e\Big[\binom{n_1-1}{i}\binom{n_2}{e}-\binom{n_2}i\binom{n_1-1}{e}\Big]{Q}^i.
  \end{align*}
It is straightforward to check that with $n_1\leq n_2$ each term inside the brackets on the last line is nonnegative, and so
\begin{gather*}
    f(n_1-1) f(n_2+1)\le f(n_1)f(n_2),\\
    \frac{q^{n_1}}{f(n_1)}\cdot\frac{q^{n_2}}{f(n_2)}\le \frac {q^{n_1-1}}{f(n_1-1)}\cdot \frac{q^{n_2+1}}{f(n_2+1)}.
       %\frac {q^{n_1+n_2}}{f(n_1)f(n_2)}\le \frac {q^{n_1-1}}{f(n_1-1)}\cdot \frac{q^{n_2+1}}{f(n_2+1)}.
\end{gather*}
\end{IEEEproof}

Similar (but simpler) checks can be performed to verify the log-convexity of the Plotkin and Singleton bounds, and we obtain the following corollary.
 
\vspace*{.1in}
\begin{corollary} \label{cor:MB}
Let $\cC$ be an $(n,k,r,\rho)$ $q$-ary LRC code with distance $d$, and let $\mu$ be defined in \eqref{eq:mu}. The following bounds hold true:
\begin{enumerate}
  \item 
 Locality-dependent Hamming bound:
   \begin{equation}\label{eq:H}
   k\leq {\mu}
    \Big({N}- \log_q \Big(\sum_{e=0}^{\lfloor \frac{\rho-1}{2}\rfloor}\binom{{N}}{e}{\color{black}(q-1)^e}\Big)\Big).
    \end{equation}
  \item
    Locality-dependent Plotkin bound: Let  $\rho>\textstyle{\frac{q-1}q}{N},$ then
     \begin{equation}
      k \leq \mu\log_q \frac{\rho }{\rho -\frac{q-1}q{N}} \label{eq:P}.
    \end{equation}
  \item
     Locality-dependent Singleton bound:
    \begin{equation}
    k\leq {\mu}r .\label{eq:S}
    \end{equation}
\end{enumerate}
%  \begin{align}
% \text {Locality-dependent sphere-packing bound:}\nonumber \\
%   & \hspace*{.0in}k\leq {\mu}
%    \Big({N}- \log_q \Big(\sum_{e=0}^{\lfloor \frac{\rho-1}{2}\rfloor}\binom{{N}}{e}{\color{black}(q-1)^e}\Big)\Big)\label{eq:H}\\[-.05in]
%\text{Locality-dependent Plotkin bound:} \nonumber \\
%&\hspace*{.3in}\text{Let  $\rho>{N}\textstyle{\frac{q-1}q},$ then}\nonumber\\
%   & \hspace*{.6in}k\leq {\mu}
%\log_q \frac{\rho }{\rho -\frac{q-1}q{N}} \label{eq:P}
%\\\text{ Locality-dependent Singleton bound:}\nonumber \\
%   &\hspace*{.6in} k\leq {\mu}r .\label{eq:S}
%  %    \end{aligned}
%  \end{align}
\end{corollary}

The bound \eqref{eq:S} is slightly weaker than the Singleton-type bounds in~\eqref{eq:msft} and~\eqref{eq:rho}. 
%previously derived in \cite{Gop11} for $\rho=2$ and in \cite{PKLK2012} for arbitrary $\rho\ge 2$. 
%\blue{\sout{The Hamming bound is tighter than the Singleton bound for $\rho\ge 3$ and is inferior to it for $\rho=2.$} (Hu: I think this argument is not always true.)}

\begin{remark}
Not all bounds on codes are log-convex in the code length.
For example, let $M_2(l, \rho)$ be the maximum size of a binary code of length $l$ and distance $\rho.$
We have $M_2(7,4)=8, M_2(8,4)=16, M_2(9,4)=20,$ and so $M_2(8,4)^2>M_2(7,4)M_2(9,4),$ violating the log-convexity condition
(which stipulates that the geometric average be greater than the ``middle value").
\end{remark}

%{\color{red} (how is the portion after the remark in this section related to the current results?)}
\textcolor{black}{Let
  $$
  R_q(r,\rho,\delta)=\limsup_{n\to\infty}\frac 1n \log_q {M_q(n,r,\rho,\delta n)},
  $$
  where ${M_q(n,r,\rho,\delta n)}$ is the maximum cardinality of the $(n,k,r,\rho)$ LRC code {with minimum distance $\delta n$}.
  We finish this section by showing that {the bound \eqref{eq:S} can be combined }with a known result to
derive an exact value of $R_q$ for $\delta=0.$
The following} lower asymptotic bound for LRC codes was obtained in \cite{BTV17}. 
\vspace*{.1in}\begin{theorem} [\cite{BTV17}] {Assume that there exists a $q$-ary MDS code of length ${N}=\rho+r-1$ and distance $\rho.$ Then the following Gilbert-Varshamov type bound holds true:
  \begin{equation}\label{eq:GVrho}
   R_q(r,\rho,\delta) \ge \frac{r}{{N}}-\min_{0< s\le 1}\Big\{\frac {\log_q b_\rho(s)}{{N}}-\delta \log_q s\Big\}   
    \end{equation}
    where
    \begin{align*}
    b_\rho(s)=&1+(q-1)\sum_{w=\rho}^{{N}}\binom {{N}}w
       s^w q^{w-\rho}\\
    &\times\sum_{j=0}^{w-\rho} \binom{w-1}j(-q)^{-j}.
    \end{align*}
    }
 \end{theorem}
This result implies the following corollary, which for $q=2$ was already established in \cite{TBF16}.
\begin{corollary} { 
Assume that there exists a $q$-ary MDS code of length ${N}$ and distance $\rho,$ then $R_q(r,\rho,0)=\frac{r}{{N}}.$
}
\end{corollary}
%\vspace*{.1in}\hspace*{.2in}{\em Proof:}
\begin{IEEEproof}
The bound \eqref{eq:S} implies the estimate $R_q(r,\rho,0)\le\frac{r}{{N}}$ while \eqref{eq:GVrho} gives
the opposite inequality.
\end{IEEEproof}
%\hfill\qed

\section{Algebraic Combinatorics of LRC Codes and LP Bounds}\label{sect:LP}

%\subsection{Preliminaries}
Delsarte's linear programming bound is a powerful method of estimating the size of optimal codes in various metric spaces
that satisfy a set of general assumptions \cite{del73}. In this section we develop an adaptation of the approach 
in~\cite{del73} to $(n,k,r,\rho)$ LRC codes.

%{\color{red} (This section is all preliminaries. There is no other subsection. Why is this subsection created ?)}
\subsection{Association Schemes and their Powers}\label{sec:prel}
\subsubsection{Metric Association Schemes}
We begin with a brief reminder about metric association schemes \cite{del73,bro89}.
Let $X$ be a finite metric space with distance function $d$, and let
$\mathbf{R}=\{R_0,R_1,\ldots, R_n\}$ be a partition of $X\times X$ such that $R_i:=\{(x,y)\in X^2 \mid d(x,y)=i\}$ for all $i$. The pair $\cA=(X,\mathbf{R})$
is called an {\em association scheme} if the intersection volume of two balls in $X$ depends only on the distance between their centers and the radii of the balls.
For each $i$ denote by $A_i$ the $|X|\times|X|$ adjacency matrix of $R_i$, where $(A_i)_{x,y}=1$ if $(x,y)\in R_i$ and $0$ otherwise. 
%Let $\mathcal{A}=\{A_i\}_{i=0}^n$. We call $(X,\mathcal{A})$ (or $(X,\{R_i\}_{i=0}^n)$) a {\it $d$-class association scheme} if there exist numbers $p_{i,j}^k$ such that
%\[
%A_iA_j=\sum_{k=0}^dp_{i,j}^kA_k.
%\]
%These numbers are called the intersection numbers of the scheme. 
The matrices $A_0,A_1,\ldots,A_n$ span a complex semisimple  algebra of dimension $n+1$, called the {\it Bose-Mesner algebra} of the scheme. 
%With respect to the basis $A_0,A_1,\cdots,A_d$, the matrix of the multiplication by $A_i$ is denoted by $B_i$, namely
%\[
%A_i(A_0,A_1,\cdots,A_d)=(A_0,A_1,\cdots,A_d)B_i,\;0\leq i\leq d.
%\]
Since each $A_i$ is symmetric, this algebra is commutative. It affords a dual basis of minimal idempotents $E_0,E_1,\ldots,E_n.$
\textcolor{black}{We can represent the matrix $A_i$ as a linear combination of the idempotents. The coefficients of this expansion form the
  \emph{first eigenmatrix} of the scheme $\cA$, denoted by $P$. A similar transition can be performed in the other direction, and the corresponding 
  coefficients form the \emph{second eigenmatrix} of $\cA,$ denoted by $Q$. Namely, we have}
   \begin{align*}
   A_i&=\sum_{j=0}^n P_{ji}E_j, \quad 0\le i\le n,\\
    E_j&=\frac1{|X|}\sum_{i=0}^n Q_{ij}A_i, \quad 0\le j\le n.
   \end{align*}
%    are called the first and the second eigenmatrices of $\cA.$
%    {\color{red} define $P_{ij}$, $Q_{ij}$}
%The $(d+1)\times (d+1)$ matrix $P$ such that
%\[
%(A_0,A_1,\cdots,A_d)=(E_0,E_1,\cdots,E_d)P
%\]
%is called the {\it first eigenmatrix} of the scheme. Dually, the $(d+1)\times (d+1)$ matrix $Q$ such that
%\[
%(E_0,E_1,\cdots,E_d)=\frac{1}{|X|}(A_0,A_1,\cdots,A_d)Q
%\]
%is called the {\it second eigenmatrix} of the scheme. Clearly, we  have $PQ=|X|I$.

\subsubsection{Products of Association Schemes}
Let $\cA=(X,\mathbf{R})$ be a metric association scheme with eigenmatrices $P$ and $Q$, and let $Y:=X^s$ be a Cartesian power of $X$.
We can define a product association scheme $\cA^{\otimes s}=(X,\mathbf{R})^{\otimes s}$ 
%Delsarte extension of $\cA$ (alternatively known as the product scheme \cite{BHPS1998,Ma1999}) 
by introducing the relations $R_{{\bfi}}, {\bfi}=(i_1,\dots,i_s)$ on $Y\times Y$
in the following obvious way \cite[p.17]{del73}:
  \begin{align*}
  R_{{\bfi}}=\{((y_{1},\dots,y_{s}),&(y'_{1},\dots,y'_{s}))\mid\\
         &(y_{j},y'_{j})\in R_{i_j}, j=1,\dots,s\}.
   \end{align*}     
The adjacency matrices of $\cA^{\otimes s}$ are formed of the Kronecker products $\otimes_{i=1}^s A_i(=A^{\otimes s}),$ where $A_i(=A)$ is an adjacency
matrix of the $i$th copy of $X$ in the product.
%Here we describe the product of association schemes. refer to  for more
%details.
%For $1\le i\le s$, let $(Y_i,\mathcal{A}_i)$ be a $d_i$-class association scheme with adjacency matrix
%$\mathcal{A}_i$. The \emph{direct product} of these schemes is the association scheme
%$$
%  (X,\mathcal{A})=(Y_1,\mathcal{A}_1)\otimes\cdots \otimes (Y_s,\mathcal{A}_s)
%$$
%defined by
%$$
%  X = Y_1\otimes Y_2\otimes \cdots \otimes Y_s
%$$
%and
%$$
%\mathcal{A}=\{\otimes_{i=1}^sM_i: M_i\in\mathcal{A}_i, 1\le i \le s\}
%$$
%where
%$$
%  \otimes_{i=1}^sM_i = M_1\otimes M_2\otimes \cdots \otimes M_s
%$$
%is the $m$-fold Kronecker product of matrices. 
%The fact that this always gives an association scheme
%follows from the properties
%\begin{align*}
%  (\otimes_{i=1}^s M_i)(\otimes_{i=1}^s N_i) &= \otimes_{i=1}^s(M_iN_i),\\
%  (\otimes_{i=1}^s M_i)\circ (\otimes_{i=1}^s N_i) &= \otimes_{i=1}^s(M_i\circ N_i),
%\end{align*}
%of the Kronecker product.  From the first identity, it follows that 
It is not hard to check that the first (second) eigenmatrix of the scheme $\cA^{\otimes s}$ equals the $s$th Kronecker power of $P$ (resp., of $Q$).
%product scheme is given by the Kronecker product of the first (resp. second) eigenmatrices for the component
%schemes $(Y_i, \mathcal{A}_i)$.
%Similarly, the second eigenmatrix $Q$ for the product scheme
%is given by $\otimes_{i=1}^s Q_i$ where $Q_i$ is the second eigenmatrix for the $i$-th component scheme.

\subsubsection{The Linear Programming Bound}
Let $(X,\cA)$ be an association scheme with $n$ classes and let $\cC$ be a code (any subset $\cC\subset X$).
The distance distribution of $\cC$ is given by $\bfa=(a_0,a_1,\dots,a_n),$ where $a_i=|(\cC\times \cC)\cap R_i|/|\cC|$ is
the average number of codewords at distance $i$ from a given codeword of $\cC$.
Clearly, $a_0=1$ and $\sum a_i = |\cC|$.
The vector $\bfa Q,$ called the {\em MacWilliams transform} of the distance distribution of $\cC$,
satisfies the Delsarte inequalities $(\bfa Q)_i \ge 0, i=1,\dots,n$. This gives rise to Delsarte's \emph{linear programming bound} on codes:
let $\cC\subset X$ be a code with distance $d$, then
\begin{align*}
  |\cC| \le \max \Big\{ &\sum_{i=0}^n a_i  \text{ s.t. } \bfa Q\ge 0,  a_0=1,\\ 
       &\hspace{-.9cm}a_{i}=0 \text{ for } 1\le i\le d-1, a_i\ge0 \text{ for } d\le i\le n \Big\}.
\end{align*}
(e.g., \cite[Ch.2,3]{del73}, \cite[Ch.17]{MaSl1977}). This bound also applies to product schemes. Indeed, let $\cA^{\otimes s}=(X,\cA)^{\otimes s}$ and let $\cC\subset X^s$ be a code.
Let $\bfa=(a_{\underline i}),$ where ${\bfi}=(i_1,\dots,i_s)$ and $i_j=0,\dots,n$ for all $j,$ 
be the distance distribution of $\cC$. Similarly, $a_{\underline 0}=1$ and $\sum a_{\bfi} = |\cC|$.
Suppose that $a_{\bfi}=0$ if $\bfi\not\in \{\underline{0}\}\cup T$ where $T$ is some subset of $\{0,\dots,n\}^s\backslash\{\underline{0}\}$. Then we have 
   \begin{align}\label{eq:lp}
     \begin{split}
     |\cC|\le\max\Big\{  &\sum_{{\bfi}\in\{0,\dots,n\}^s} a_{{\bfi}} \text{ s.t. }\\
      &\hspace{0.5cm} \sum_{{\bfi}\in\{0,\dots,n\}^s} a_{{\bfi}}Q_{{{\bfi}},{{\bfj}}}\ge0 \text{ for all } \bfj;\\
      &\hspace{0.5cm} a_{\underline 0}=1, a_{\bfi}=0 \text{ for } \bfi\not\in \{\underline{0}\}\cup T,\\
   &\hspace{0.5cm} a_{{\bfi}}\ge 0 \text{ for }{\bfi} \in T\Big\}
     \end{split}
   \end{align}
where $Q$ is the second eigenmatrix of $\cA^{\otimes s}.$ More details about product schemes are given in 
\cite[Sec. 2.5]{del73} as well as in more recent works \cite{BHPS1998,Ma1999}.

\subsubsection{The Hamming Scheme}
The following classic example will be useful below in the context of LRC codes.
Let $F$ be a set of cardinality $q\ (q\ge 2)$ and let $X=F^n$ be a Cartesian power of $F$. 
We specialize the definition of the metric scheme by assuming that $d$ is a Hamming metric on $X$.
Namely, let $R_i:=\{(x,y)\in X^2\mid d(x,y)=i\}$ where $d$ is the Hamming distance.
We obtain a symmetric association scheme with $n$ classes, denoted by
$H(n,q)$. The eigenvalues of $H(n,q)$ are given by $Q_{ij}=K^{(n)}_j(i)$ \cite{del73},
where
$$
K_j^{(n)}(x)=\sum_{l=0}^j(-1)^l(q-1)^{j-l}\binom x l\binom {n-x}{j-l}
$$
is the \emph{Krawtchouk polynomial}. Also we have $P=Q$.

The Hamming scheme $H(n,q)$ also carries the structure of a product scheme for $n\ge2$.
Consider the Hamming scheme $H(m+n,q)$ as being obtained from the product of $H(m,q)$ and $H(n,q)$ 
by merging all relations $R_{i',i''}$ with $i'+i''=i$ into one relation $R_i$.  We have 
$A_i = \sum_{i'+i''=i}A_{i'i''}, E_i = \sum_{i'+i''=i}E_{i'i''}$ and 
$Q_{ij} = \sum_{j'+j''=j} Q_{i'j'}Q_{i''j''}$ for any pair $(i',i'')$ with $i'+i''=i.$ 
Also we can view all three association schemes involved as merged versions of powers of $H(1,q)$.

Clearly, $H(n,q)=(H(1,q))^{\otimes n},$  and similarly $H(st,q)=(H(t,q))^{\otimes s}$ for any $s\ge 2.$
 We conclude that the eigenvalues of the scheme $H(st,q)$ have
the form
   \begin{equation}\label{eq:K}
     P_{\bfi,\bfj}=Q_{\bfi,\bfj}=K_{{\bfj}}^{(t)}({\bfi})=\prod_{p=1}^s K_{j_p}^{(t)}(i_p),
  \end{equation}
  where the multi-indices $\bfi,\bfj$ are the indices of the relations of the scheme. As is the case
  with the original Hamming scheme, the obtained scheme $H(st,q)$ is also self-dual.
It is this setting that we apply to the analysis of LRC codes in the next section.

\subsection{The Linear Programming Bound for LRC Codes with Disjoint Repair Groups}
\subsubsection{General Bound}
We begin with stating a general LP bound for codes with locality.
   Let $\cC$ be an $(n,k,r,\rho)$-LRC code with minimum distance $d$.
Suppose that $n=s{N}$ where $N=r+\rho-1$. For $0\le t\le s-1$, define the interval 
   $$
   \cR_{t+1}=[t{N}+1,(t+1){N}],
   $$
   so the coordinate set is a disjoint union of these intervals:
   $$
   [n]=\cup_{i=1}^s \cR_i.
   $$
For $I\subset [n]$ denote by $\cC\vert_{I}$ the projection of $\cC$ on the coordinates in $I$.
Throughout this section we will assume that the code $\cC$ has the property that
  $$
     d(\cC\vert_{\cR_i})\ge \rho \text{ for all } i=1,\dots,s.
  $$
In accordance with \eqref{eq:K}, define the following polynomials of $s$ discrete variables
$\bfx=(x_1,\ldots,x_s)$
$$
K^{({N})}_{\bfj}(\bfx)=\prod_{p=1}^sK^{({N})}_{j_p}(x_p),
$$
where $\bfj=(j_1,\ldots,j_s).$  
The polynomials $K^{({N})}_{\bfj}$ are orthogonal on the set $\{0,\dots,N\}^s$:
  $$
    \sum_{\bfi\in \{0,\dots,N\}^s} e(\bfi)K^{({N})}_{\bfj}(\bfi)K^{({N})}_{\bfj'}(\bfi)=q^n e(\bfj)\delta_{\bfj,\bfj'}
  $$
where
  \begin{equation}\label{eq:Q0}
  e(\bfi)=K_{\bfi}^{({N})}({\underline 0})=\prod_{p=1}^s \binom{{N}}{i_p}(q-1)^{i_p},
  \end{equation}
and $\delta_{\bfj,\bfj'}$ is the Kronecker delta function.

Let $\bfa=(a_{\bfi}, \bfi\in \{0,\dots,N\}^s)$ be the distance distribution of $\cC$, where each
  $  {\bfi}=({i_1},{i_2},\dots,{i_s})
  $
is an $s$-tuple. Here $a_\bfi$ is the number of pairs of codewords 
$c=(c_1,c_2,\ldots,c_s),c'=(c'_1,c'_2,\ldots,c'_s)\in \cC$ such that the Hamming distance
$d(c_{j},c'_{j})={i_j}, j=1,\dots,s,$ normalized by the cardinality of the code $q^k.$
Note that the codewords $c_{j},c'_{j}$ are contained in the code $\cC\vert_{\cR_j}.$
By definition, we have $a_{{\underline 0}}=1$, and $a_{\bfi}=0$ if $\bfi\not\in \{\underline{0}\}\cup T$ where 
$T=\{\bfi=(i_1,\ldots,i_s)\mid i_1+\cdots+i_s\ge d, i_j\in\{0,\rho,\rho+1,\dots,{N}\} \text{ for all }j=1,\dots,s\}.$

Now it is direct to check that the general bound of \eqref{eq:lp} in our case takes the following form.
\vspace*{.1in}
\begin{theorem}[Primal LP bound]\label{thm:primal}
Let $\cC$ be a $q$-ary $(n,k,r,\rho)$ LRC code with distance $d$. Define
  \begin{multline*}
   T:=\big\{\bfi=(i_1,\ldots,i_s)\mid i_1+\cdots+i_s\ge d, \\
   i_j\in\{0,\rho,\rho+1,\dots,{N}\}
     \text{ for all } j=1,\dots,s
    \big\}.
   \end{multline*}
Then the cardinality of $\cC$ satisfies $|\cC|\le 1+\sum_{\bfi\in T} a_{\bfi},$
where the vector $(a_{\bfi}, \bfi\in T)$ is a solution of the following LP problem
\begin{gather*}
  \mbox{\rm maximize }  \sum_{\bfi\in T}a_{\bfi}\\
  \mbox{\rm subject to }  a_{\bfi}\ge 0, \ {\bfi}\in T,\\
  \hspace{-.5cm} \sum_{{\bfi}\in T}a_{{\bfi}}Q_{{\bfi}{\bfj}}
  \ge-K_{{\bfj}}^{(N)}({\underline 0}),\
  {\bfj}\in\{0,\dots,N\}^s\backslash\{{\underline 0}\}.
\end{gather*}
\end{theorem}

The dual problem of the LP problem in Theorem \ref{thm:primal} has the following form.
\begin{theorem}[Dual LP bound]\label{thm:dual} Let $\cC$ and $T$ be as defined in Theorem \ref{thm:primal}. The cardinality of $\cC$ satisfies
  \begin{align}
  |\cC|\le &1+\min \sum_{\bfj\in\{0,\dots,N\}^s\backslash\{{\underline 0}\}}f_{\bfj} K_{\bfj}^{(N)}({\underline 0})\label{eq:dualLP}\\
&\mbox{\rm subject to} \nonumber\\
  &f_\bfj\ge0, \quad \bfj\in\{0,\dots,N\}^s\backslash\{{\underline 0}\}, \label{eq:cf}\\
  %& f(\bfi)\le 0, \quad \bfi\in T.\label{eq:neg}
  & 1+\sum_{\bfj\in\{0,\dots,N\}^s\backslash\{{\underline 0}\}}
  f_{\bfj}K_{\bfj}^{(N)}(\bfi) \le 0, \quad \bfi\in T.\label{eq:neg}
\end{align}
\end{theorem}
As in the classical case  (cf. \cite[p.53]{del73},\cite{MaSl1977}), instead of solving this LP problem, we construct feasible solutions which provide upper bounds for the minimum. We state the result in polynomial form, which is obvious from \eqref{eq:dualLP}-\eqref{eq:neg}.

\vspace*{.1in}
\begin{corollary}\label{cor:poly}
{\em Let $\cC$ be a $q$-ary $(n,k,r,\rho)$ LRC code with distance $d$. 
Let $f(\bfx)=f(x_1,\dots,x_s)$ be a polynomial whose Krawtchouk expansion has the form
  $$
  f(\bfx)=1+\sum_{\bfj\in\{0,\dots,N\}^s\backslash\{{\underline 0}\}}
  f_{\bfj}K_{\bfj}^{(N)}(\bfx),
  $$
  where the coefficients $f_\bfj$ satisfy (i) $f_\bfj\ge0$  for $\bfj\in\{0,\dots,N\}^s\backslash\{{\underline 0}\}$, and (ii) $f(\bfi)\le 0$ for $\bfi\in T.$ 
%the conditions in \eqref{eq:cf}-\eqref{eq:neg}.
Then $|\cC|\le f({{\underline 0}}).$
}
\end{corollary}
%\begin{IEEEproof}
%\end{IEEEproof}
\vspace*{.1in}

%% \begin{problem} 
%\begin{align*}
%  \mbox{\rm minimize } \sum_{\mathbf{j}\in[{N}]^s\backslash\{{\underline 0}\}}
%  \mathbf{b}_{\mathbf{j}}K_{\mathbf{j}}^{{N}}({\underline 0}),\
%\end{align*}
%{\rm subject to}
%\begin{align*}
%  \mathbf{b}_{\mathbf{j}} &\ge 0, \ \mathbf{j}\in[{N}]^s\backslash\{{\underline 0}\},\\
%  \sum_{\mathbf{j}\in[{N}]^s\backslash\{{\underline 0}\}}\mathbf{b}_{\mathbf{j}}Q_{\mathbf{i}\mathbf{j}} &\le -1,\ \mathbf{i}\in T.
%  \end{align*}
%%\end{problem}

\subsubsection{The Singleton Bound}
The bounds in Corollary~\ref{cor:MB} can be proved using the poly\-no\-mial approach of Corollary~\ref{cor:poly}. To exemplify this claim,
we give another proof of the Singleton bound. The original form of this bound in \cite{PKLK2012} is as follows:
\begin{align}\label{Singleton}
  d \le n-k+1-\Big(\Big\lceil \frac{k}{r} \Big\rceil -1\Big)(\rho-1).
\end{align}
Recall that $n=sN=s(r+\rho-1)$. Assume that $d=t{N}+\partial$ for some $\partial$, $1\le \partial\le {N}$.
Relaxing \eqref{Singleton} by omitting the ceiling function, we obtain
\begin{align}\label{Singleton_no_ceil}
k \le r(s-t)+\frac{r(\rho-\partial)}{{N}}.
%&\le \frac{n-d+\rho}{{N}}\cdot r \\
\end{align}

Recall that in the classical case the Singleton bound is proved using the polynomial \cite[p.~54]{del73}, \cite[p.~544]{MaSl1977}
%\begin{lemma}\cite[p.544, Example (4)]{MaSl1977}\label{lemma:decomposition}
  \begin{align}
  f(x) &= q^{n-d+1}\prod_{j=d}^{n}\Big(1-\frac{x}{j}\Big) \nonumber\\
  &= \sum_{k=0}^{n}\binom{n-k}{d-1}K_k^{(n)}(x)\Big/ \binom n{d-1} \label{eq:nn}
  \end{align}
  (the ``annihilator'' of the weight distribution).
%\end{lemma}
Following this approach,
define the polynomial $f(\bfx)$ in the form
%\begin{figure*}
 \begin{equation}\label{polynomial}
       f(\bfx) = \left\{
    \begin{aligned}
      &\prod_{j=1}^{s-t-1}q^{r}\prod_{i=\rho}^{{N}}\Big(1-\frac{x_j}{i}\Big) 
      %\\ &\hspace*{.2in} \times 
      q^{r+\rho-\partial}\prod_{i=\partial}^{{N}}\Big(1-\frac{x_{s-t}}{i}\Big)\\
        &\hspace{5cm} \mbox{if } \rho \le \partial \\
      &\prod_{j=1}^{s-t} q^{r}\prod_{i=\rho}^{{N}}\Big(1-\frac{x_j}{i}\Big) 
       \hspace{2cm}\mbox{if } \rho > \partial.
    \end{aligned}
    \right .
  \end{equation}
%\end{figure*}
We will prove that the polynomial $f$ is a feasible solution of the dual LP problem, i.e., that 
it satisfies the conditions in \eqref{eq:cf}, \eqref{eq:neg}. Consider the expansion
    $$
 f(\bfx)=1+\sum_{\bfj\in\{0,\dots,N\}^s\backslash\{{\underline 0}\}}f_{\bfj}K^{({N})}_{\bfj}(\bfx).
   $$
On account of \eqref{eq:nn} we conclude that $f_{\bfj}\ge0$ for all $\bfj$, so \eqref{eq:cf} is indeed true.

To prove \eqref{eq:neg}, we will show that 
$f(\bfi)\le 0$ for $\bfi\in T$.
First suppose that $\rho\le \partial$. Choose any $\bfi=(i_1,\dots,i_s)\in T$, then $i_1+\cdots+i_s\ge d$. If $i_{s-t}+\cdots+i_s\ge d$, then we must
have $i_{s-t}\ge \partial$, implying that $f(\bfi)=0$. If $i_{s-t}+\cdots+i_s < d$, then there must
exist some nonzero $i_l, 1\le l\le s-t-1$, which implies that $i_l\ge\rho$ and again $f(\bfi)=0.$
The case $\rho>\partial$ can be analyzed using similar arguments.

Therefore, by Theorem \ref{thm:dual} we obtain the bound
\begin{align*}
  |\cC| \le f({\underline 0})&=
  \begin{cases}
    q^{r(s-t)+\rho-\partial} & \mbox{if } \rho \le \partial \\
    q^{r(s-t)}        & \mbox{if } \rho >   \partial,
  \end{cases}
\end{align*}
In other words,
\begin{align*}
  k \le 
  \begin{cases}
    r(s-t)+\rho-\partial & \mbox{if } \rho \le \partial \\
    r(s-t)        & \mbox{if } \rho >   \partial.
  \end{cases}
\end{align*}
This estimate is an LP version of the Singleton bound, and it is slightly better than~\eqref{Singleton_no_ceil}.

\subsubsection{Bounds for $(n,k,r,2)$ LRC Codes}\label{sect:LPfinite}
It is interesting to apply the LP approach to bounds on LRC codes for $\rho=2$, i.e., the case of single-symbol locality (Definition \ref{def1}).
 The known bounds that apply in this case include the Singleton bound \eqref{eq:S}, which does not
depend on $q$, and a shortening bound of \cite{CaMa2015}. 
For the ease of reading we reproduce the bound \eqref{eq:cm}:
{\em For any $(n,k,r,2)$ LRC code with distance $d$, we have
  $$ 
    k\le \min_{1\le s\le n/(r+1)} \,\{sr+\log_q M_q(n-s(r+1),d)\},
    %  k \le \min_{t\in\integers_{\ge0}}\{tr+\log_q A_q(n-t(r+1),d)\}
  $$
 where $M_q(n,d)$ is the maximum cardinality
of a $q$-ary code of length $n$ and distance $d.$  }
%
%For any $q$-ary LRC code with the parameters $(n,k,r)$ and distance $d,$
%  \begin{equation}\label{eq:cm}
%    k\le \min_{1\le s\le n/(r+1)} \,\{sr+\log_q M_q(n-s(r+1),d)\},
%  \end{equation}
%where $M_q(n,d)$ is the maximum cardinality of a $q$-ary length $n$ code with distance $d$.

\vspace*{.1in}We computed this bound and the bound of Theorem \ref{thm:dual}, using $\rho=2$ in the definition of the index set $T$.
The results are summarized in Tables I--IV. Note that the corresponding length of the code is $n=s(r+\rho-1)=s(r+1)$, and the entry of the table is the upper bound on the dimension $k$.
We perform the computations using the GAP package GUAVA and the package 
GLPK in the symbolic computations system SageMath \cite{sage}. 
Each result was verified using the package COIN-OR, also available in SageMath.

\begin{table}[ht] 
 \centering 
    \caption{$q=2, d=3, s=2$} 
    \begin{tabular}{|c|c|c|c|c|c|c|c|c|c|} \hline 
      r & 2   &  3  &   4   &   5   &  6 &  7 & 8 & 9  & 10\\\hline 
%      \text{SHG}  & 3.0 & 4.0 & 6.0 & 8.0 & 10.0 & 11.644 & 13.672 & 15.409 & 17.409\\ 
      \text{SH}  & 3 & 4 & 6 & 8 & 10 & 11 & 13 & 15 & 17\\ 
      \text{LP}   &2 & 4 & 5 & 7 & 9 & 11 & 12 & 14 & 16 \\\hline 
    \end{tabular} 
\vspace*{.1in}  
    \caption{$q=2, d=3, s=3$} 
    \begin{tabular}{|c|c|c|c|c|c|c|c|c|c|} 
      \hline 
      r & 2   &  3  &   4   &   5   &  6 &  7 & 8 & 9  & 10\\\hline 
      %\text{SHG}  & 5.0 & 7.0 & 10.0 & 13.0 & 16.0 & 18.644 & 21.672 & 24.409 & 27.409\\ 
      \text{SH}  & 5 & 7 & 10 & 13 & 16 & 18 & 21 & 24 & 27\\ 
      \text{LP}   &4 & 7 & 9 & 12 & 15 & 18 & 20 & 23 & 26 \\\hline 
    \end{tabular} 
 \vspace*{.1in}

    \caption{$q=2, d=5, s=2$} 
    \begin{tabular}{|c|c|c|c|c|c|c|c|c|c|} 
      \hline 
    r  & 2   &  3  &   4   &   5   &  6 &  7 & 8 & 9  & 10\\\hline 
    %  \text{SHG}  & 2 & 3 & 5.0 & 6.0 & 7.585 & 9.0 & 10.585 & 12.585 & 14.585\\ 
      \text{SH}  & 1 & 2 & 3 & 5 & 7 & 8 & 10 & 12 & 13\\ 
      \text{LP}   &1 & 2 & 3 & 5 & 6 & 8 & 9 & 11 & 13 \\\hline 
    \end{tabular} 
 \vspace*{.1in}

    \caption{$q=2, d=5, s=3$} 
    \begin{tabular}{|c|c|c|c|c|c|c|c|c|c|} \hline 
    r  & 2   &  3  &   4   &   5   &  6 &  7 & 8 & 9  & 10\\\hline 
    %  \text{SHG}  & 3 & 5 & 7.585 & 10.555 & 13.257 & 15.741 & 18.484 & 21.258 & 23.876\\ 
      \text{SH}  & 2 & 5 & 7 & 10 & 13 & 15 & 18 & 21 & 23\\ 
      \text{LP}   &2 & 4 & 6 & 9 & 11 & 14 & 17 & 19 & 22 \\\hline 
    \end{tabular} 

\end{table} 
In all the above examples, the LP bound either matches the shortening bound or is tighter than it.

\section{Asymptotic Bounds for Binary Linear LRC Codes} \label{sec:asym}
In this section we study asymptotic bounds for binary linear LRC codes that can locally correct one erasure. 
%However, the results extend to the case of general $q$-ary alphabet quite straightforwardly.
 Throughout the section we assume that the code satisfies Definition \ref{def1}. For $\delta\in[0,1/2],$ define the functions
   \begin{align*}
   R(r,\delta)&=\limsup_{n\to\infty}\frac 1n \log_2 M(n,r,\delta n)\\
   R^{(\text{\rm lin})}(r,\delta)&=\limsup_{n\to\infty}\frac 1n \log_2 M^{(\text{\rm lin})}(n,r,\delta n),
   \end{align*}
where $M(n,r,d)$ (respectively, $M^{(\text{\rm lin})}(n,r,d)$) is the maximum cardinality of a code (respectively, of a linear code) of length $n$, distance $d$ and locality $r$. Clearly, $R(r,\delta)\ge R^{(\text{\rm lin})}(r,\delta).$ The best currently known asymptotic bounds on binary LRC codes are described in the following theorem.
  \begin{theorem} [\cite{CaMa2015,TBF16}]\label{thm:literature} We have    
    \begin{align}\label{eq:GV}
    \begin{split}
    R^{(\text{\rm lin})}(r,\delta) \ge 1-&\min_{0<s\le 1}\Big\{\frac 1{r+1}\log_2((1+s)^{r+1} \\
     &+(1-s)^{r+1})-\delta\log_2s\Big\},
     \end{split}
    \end{align}
  \begin{align} 
  \begin{split}
    R(r,\delta) \leq &\min_{0 \leq \sigma < 1/(r+1)} \Big\{ \sigma r \\
    &+ (1-\sigma(r+1)) R_\text{\rm opt}\Big(\frac{\delta}{1-\sigma(r+1)}\Big)\Big\}, \label{eq:upper}
    \end{split}
  \end{align}
where $R_\text{\rm opt}(\delta)$ is any asymptotic upper bound on the rate of codes with relative distance $\delta$.
These bounds imply that
  \begin{gather*}
    R^{(\text{\rm lin})}(r,\delta)>0 \text{ iff } 0\le \delta<1/2; \\
    R(r,0)=R^{(\text{\rm lin})}(r,0)=\frac r{r+1}.
    \end{gather*}
   \end{theorem}
  The lower bound \eqref{eq:GV} is of the Gilbert-Varshamov type and was derived in \cite{CaMa2015} and  \cite{TBF16}, while the bound \eqref{eq:upper} is obtained from \eqref{eq:cm} by passing to the limit of large block length $n$ (see \cite{CaMa2015}).
To obtain the tightest possible bound in \eqref{eq:upper} we substitute the best known bound on $R_\text{\rm opt}(\delta)$, i.e., the McEliece et al. bound \cite{mce77}:
    \begin{equation}\label{eq:mrrw2}
   R_\text{\rm opt}(\delta)\le 1+ \min_{0<\alpha\le 1-2\delta}( g(\alpha^2) - g(\alpha^2 +2\delta\alpha+2\delta))\,,
    \end{equation}
where
$g(x): =h(\frac12-\frac12\sqrt{1-x})$ and $h(x) := -x\log_2 x -(1-x) \log_2 (1-x)$ is the binary entropy function.
  
\begin{remark} Even though in Sec.~\ref{sect:LPfinite} we showed by example that the LP bound is better than the bound \eqref{eq:cm} for finite length, it is difficult to derive a closed-form asymptotic version of the LP bound. The problem occurs because to derive the asymptotic version of the LP bound it would be easier to have a small number of local codes whose distance $\rho$ grows in proportion to $n$. In reality we have to deal with a growing number of local codes with distance $\rho=2$.
\end{remark}

%Here we will prove the following theorem which improves upon the bound \eqref{eq:upper} in the case of linear codes that are also locally repairable.
In this part we prove a new bound on linear LRC codes which which improves upon the (linear case of the) bound \eqref{eq:upper}.
We begin with a remark that for linear LRC codes the recovery functions defined in \eqref{eq:def1} are also linear.

\begin{lemma}\label{lemma:linear} Let $\cC$ be a linear LRC code of length $n$, dimension $k$, and locality $r$ over a field $\F.$ Then for every coordinate $i\in[n]$ the recovery function $\phi_i$ defined in \eqref{eq:def1} is also linear.
\end{lemma}
\begin{IEEEproof} %The code $\cC$ is the image of a linear map $\F^k\to\F^n.$ 
Choose a generator matrix $G$ of $\cC$ and let $V\subset \F^k$ be the set of columns of $G$. 
Given $x\in\F^k$, the coordinates of the codeword $c=xG$ are equal to $(x,v)$, where $v\in V$ and $(\cdot,\cdot)$ is the dot product. 
Since $\cC$ has locality $r$, the coordinate $(x,v_i)$ must be a function of the coordinates in its recovery set $\cR$. Let 
$\{v_j, j\in\cR\}$ be the corresponding subset of columns of $G$.
Without loss of generality we may assume that the $\rk(\{v_j,j\in \cR\})\le k-1.$

If $v_i\in(\spn\{v_j,j\in \cR\}),$ then there exists a linear recovery function for the $i$th coordinate, so let us
assume the opposite (thus, $v_i\ne 0$). Let $x\in \F^k$ be such that $(x,v_i)\ne 0$ and $(x,v_j)=0, j\in \cR.$
Further, let $y\in \F^k$ be an arbitrary vector and consider the codewords $yG$ and $(x+y)G.$ Clearly, their entries in the
coordinates of $\cR$ are the same, so the LRC property implies that their $i$th coordinates are equal as well. This however is
clearly not the case, so we obtain a contradiction.
\end{IEEEproof}

\vspace*{.1in}The main result of this subsection is given in the following theorem.
\begin{theorem}[Linear LRC codes] \label{thm:linear} The maximum rate of a linear LRC code satisfies the following inequality. 
    \begin{align}\label{eq:ldpc}
    \begin{split}
    R^{(\text{\rm lin})}(r,\delta) \leq & \min_{0 \leq s < 1/(r+1)} \Big\{s r \\
    &+ (1-s(r+1)) R_0\Big(r,\frac{\delta}{1-s(r+1)}\Big)\Big\} ,
    \end{split}
   \end{align}
where
\begin{gather}
R_0(r,\delta) := h(\tau) - c(r+1,\tau),\label{eq:R0}\\ c(w,\tau):= \frac{\log_2{e}}{8w^2}\Big(\frac{\tau^w}{2}\Big)^{w+1}, \notag\\[.05in]
\tau:=1/2-\sqrt{\delta(1-\delta)}.\label{eq:tau}
\end{gather}
\end{theorem}

Bound \eqref{eq:ldpc} improves upon the bound in \eqref{eq:upper} for all values of relative distance, however the improvement is rather mild, and can barely be seen in a plot. 

The proof of this theorem consists of two steps. First we observe that the approach of \cite{iceland2015coset} applies to LRC codes, yielding a bound on their rate. In the second step we combine this approach with a recursive shortening bound approach of \cite{CaMa2015}, obtaining Theorem~\ref{thm:slrc} below. 

To proceed with the first step, let us quote the main technical lemma of \cite{iceland2015coset}.
\begin{lemma} [\cite{iceland2015coset}] \label{lemma:ldpc}  Consider a sequence of LDPC codes with increasing length $n$ and parities of weight at most $w$. Suppose that
   the distance of codes converges to the value $\delta$ as $n\to\infty.$ Then the maximum achievable rate of the codes
   is bounded above by $R_0(w-1,\delta)$ given in \eqref{eq:R0}.
\end{lemma}

Below in Sec.~\ref{sect:is} we provide some details of the argument of \cite{iceland2015coset} because it is be needed for our second result in this section, namely a bound on linear LRC codes with disjoint repair groups. In particular, it will be clear that Lemma \ref{lemma:ldpc} also applies to linear LRC codes since the conditions required for it to hold are actually weaker than that 
the assumptions in the proof in \cite{iceland2015coset}. The weaker set of assumptions used below is as follows:
   \begin{subequations}\label{eq:union}
   \begin{align}
   \begin{split}
     \exists\{v_1,\dots,v_m\}\subset \cC^\bot \text{ s. t. } \wt(v_i)\le w, i=1,\dots,m
     \end{split}
   \end{align}
   \begin{equation}
    \bigcup_{i=1}^m \supp(v_i)=[n].
   \end{equation}
   \end{subequations}
For linear LRC codes, these conditions are satisfied for $w = r+1$. The statement in \cite{iceland2015coset} in addition to \eqref{eq:union} assumes that $\set{v_1,\dots,v_m}$ form a basis for the code $\cC^\bot$. 
It is true that this condition holds for LDPC codes, but this is an artifact of the LDPC setting rather than an essential element of the proof that we give in Sec.~\ref{sect:is}. From our discussion below (see the remarks after
Eq.~\eqref{coset_leader_constraint})
%following Proposition~\ref{lemma:pi} and the proof of  Lemma~\ref{iceland_lemma22})
 it will become clear that conditions \eqref{eq:union} suffice to complete the argument. 
As a consequence, we obtain the following bound on the rate of linear LRC codes:
 \begin{equation}\label{eq:lin}
   R^{(\text{\rm lin})}(r,\delta)\le R_0(r,\delta).
  \end{equation}

%The assumptions on the generator matrix of $\cC$ in \eqref{eq:union} are the only requirements in \Cref{iceland_lemma22,coset_leader_constraint} in \Cref{sect:is}. We in particular refer the reader to the discussion following Proposition~\ref{lemma:pi} and the proof of Proposition~\ref{iceland_lemma22}. 
  
This bound does not improve on \eqref{eq:upper}, but it is possible to establish a recursion that will lead to an improvement. Namely, we combine \eqref{eq:lin} with code shortening to obtain \eqref{eq:ldpc}. 
The following statement is a minor modification of the result of \cite{CaMa2015}.
\begin{theorem} \label{thm:slrc} Let $\cC$ be a binary  LRC code of length $n$, locality $r$ and distance $d$, then the dimension of $\cC$, $\dim(\cC)$, satisfies,
   \begin{equation}\label{eq:cmlrc}
   \dim(\cC) \le \min_{1 \leq s \leq n/(r+1)}(sr+\log_2 M(n-s(r+1),d,r))
   \end{equation}
where $M(m,d,r)$ is the maximum cardinality of a linear LRC code\footnote{The statement in \cite{CaMa2015} does not include the 
LRC condition of the shortened code.} of length $m$, distance $d$, and locality $r$.
Therefore, 
		\begin{equation}\label{eq:sa}
		\begin{split}
		R^{(\text{\rm lin})}(r,\delta) \leq &\min_{0\leq \sigma\leq 1/(r+1)} \Big\{\sigma r \\
		&+ (1-\sigma(r+1)) R^{(\text{\rm lin})}\Big(r,\frac{\delta}{1-\sigma(r+1)}\Big)\Big\}.
		\end{split}
		\end{equation}
\end{theorem}
 \begin{IEEEproof} Eq.~\eqref{eq:sa} is an obvious consequence of \eqref{eq:cmlrc}, so let us prove \eqref{eq:cmlrc}.  The proof in \cite{CaMa2015} relies on the fact that for any $s, 1\le s\le k/r$ there exists a subset of 
 coordinates $I\subset [n], |I|=s(r+1)$ such that $\log|C_I|\le sr$ (\cite[Lemma 1]{CaMa2015}). That such a subset exists can be shown relying
 on the locality property of the code $\cC.$ Let $I^c\defeq [n]\setminus I$. We shorten the code $\cC$ to obtain a code $\cC_{I^c}$ of length $n-s(r+1),$ dimension at least $k-sr,$ and distance $d$ (\cite[Lemma 2]{CaMa2015}). This code is obtained by taking all the codewords
 that contain zeros in the coordinates in $I$ and discarding these coordinates.

The only added element in our claim is that the shortened code $\cC_{I^c}$ itself is LRC. Indeed, let $i\in I^c.$
Referring to Def.~\ref{def1}, we need to prove that for any coordinate $i\in [n]\backslash I$ there exists a function $\phi_i$ that 
depends on at most $r$ other coordinates and computes the value of the $i$ th coordinate of the codeword $c\in \cC_{I^c}.$ There are two cases: 

(i) The repair group $\cR_i$ of $i$ does not intersect the subset $I$. In this case there is nothing to prove.

(ii) Some number of the coordinates of $\cR_i$ are inside the subset $I$. Let $J_i:=\cR_i\cap I.$
In this case the value of the discarded coordinates for every codeword of $\cC_{I^c}$ is equal to $0$. Suppose that 
$\psi_i(\{c_{j}, j\in \cR_i\backslash i\})$ is the recovery function of the original code $\cC.$ We claim that the recovery group of the
coordinate $i\in I^c$ in the code $\cC_{I^c}$ is the subset $\cR_i\backslash J_i$, and the recovery function is obtained from $\phi_i$
by substituting zeros for all the arguments in $J_i.$ Note that the function $\phi_i$ essentially depends on $|\cR_i|-|J_i|-1\le r$ coordinates of the codeword $c$, conforming with the locality requirement. 
 \end{IEEEproof}

\begin{remark} While we need this result only for linear codes, the  claims of Theorem \ref{thm:slrc} are still valid if we omit the linearity assumption (with obvious modifications to the statement). 
\end{remark}
  
Now Theorem \ref{thm:linear} follows immediately by using \eqref{eq:lin} in the estimate \eqref{eq:sa}.

In the next subsection we give a sketch of the approach in \cite{iceland2015coset} for LDPC codes. In Sec.~\ref{sec:DRG}, we improve on Theorem \ref{thm:linear} for the case of disjoint repair groups.

 \subsection{The Approach of  Iceland and Samorodnitsky \cite{iceland2015coset} to Bounds on LDPC Codes}\label{sect:is}
Coset graphs of linear codes have been often used as a tool to study combinatorial properties of codes and to obtain bounds on their parameters \cite{Cal85,del91,friedman2005generalized}. Given a linear code $\cC,$ define a graph $\Gamma(V,E)$, where
$V={\F}_q^n/\cC$, i.e., the vertices of $\Gamma$ 
correspond to the cosets of the code, and two cosets are connected by an edge
if the Hamming distance between them is one.

%	It is known \cite[p.136]{Cal85} that the eigenvalues $\lambda_1 \geq \lambda_2 \geq ...$ of the adjacency matrix of $\bb{T}(\cC)$ satisfies,
%	$$\lambda_i = n-2 d_i$$
%	where $d_1 \leq d_2 \leq ...$ are the weights of the codewords in $\cC$. Since $d_1 = 0$ for any linear code, we have $\delta = \frac{1}{2}(1- \lambda_2/n),$ i.e., 
%the minimum distance of the code corresponds to the second largest eigenvalue of the coset leader graph, $\bb{T}(\cC)$. 	

%{\color{red} (should we describe coset leader graphs?)}
Let $\cC$ be a linear code. Throughout this section we use the coset graph of the {\em dual code} $\cC^\bot,$ so all the references
to the coset graph below are with respect to $\cC^\bot.$ The length of the shortest path between a pair of vertices equals the Hamming distance between the corresponding cosets. Given a vertex $v\in V,$ denote by $\cB_\Gamma(v,t)$ the ball of radius $t$ around it in the graph. Since the graph is vertex-transitive,
the volume of the ball does not depend on $v$, and we will use the notation $B(t):=|\cB_\Gamma(v,t)|,$ where $v$ is an arbitrary vertex. Clearly, $B(t)$ equals the number of cosets whose leaders are of weight at most $t.$

The starting point of the argument in \cite{iceland2015coset} is the following result from \cite{friedman2005generalized}.
\begin{theorem}[\cite{friedman2005generalized}] Consider a linear codes $\cC$ of length $n$ and let $B(t)$ be the number of cosets of weight at most $t$ in $\cC^\bot.$ Then
  \begin{equation}\label{eq:FT}
  |\cC|\le 2^{o(n)} B(\tau n), %\quad\text{where $\tau=\frac12-\sqrt{\delta(1-\delta)}.$}
  \end{equation}
  where $\tau$ is defined in \eqref{eq:tau} and $\delta$ denotes the relative distance of the code $\cC$.
\end{theorem}
Using the obvious estimate $B(t)\le \sum_{i=0}^t \binom ni, t=\tau n,$ one obtains a bound valid for any code $\cC.$ The main idea in \cite{iceland2015coset} is that it is possible to obtain a tighter estimate for $B(t)$ in the case when $\cC$ is an LDPC code,
leading to an improved bound on the rate of such codes compared to the universal bounds of \cite{mce77}.

\begin{proposition}[\cite{iceland2015coset}] Let $x\in\{0,1\}^n$ be a random vector with independent Bernoulli coordinates $x_i$ such that
$P(x_i=1)=p, P(x_i=0)=1-p, p<1/2,$ and let $\pi_p$ be the probability that $x$ is a coset leader of
$\cC^\bot.$ Then
    \begin{equation}\label{eq:BB}
      B(pn)\le \pi_p \Big(\sqrt{2n}  \sum_{i=0}^t \binom ni\Big).
   \end{equation}
   \label{lemma:pi}
\end{proposition}
\begin{IEEEproof} We include a very short proof. Limiting ourselves to the vectors $x$ with at most $pn$ ones, we have
  \begin{align*}
  \pi_p &\ge B(pn) p^{pn}(1-p)^{(1-p)n}\\
  &=B(pn)2^{-h(p)n}\\
  &\ge B(pn)\frac 1{\sqrt{2n}\sum_{i=0}^t \binom ni}.
  \end{align*}
\end{IEEEproof}
The next step, which is the main technical ingredient of the result in \cite{iceland2015coset}, is to show that $\pi_p$ is an exponentially declining function of $n$. Let us assume that the dual code $\cC^\bot$ contains 
a set vectors $v_1,\dots,v_m$ such that $\wt(v_i)=w$ for all $i$ and that $\cup_{i}\supp(v_i)=[n]$.

Construct a partition $\cI_w$ of the coordinate set $[n]$ into $w$ disjoint sets $I_w,I_{w-1},\dots,I_1$ as follows. Suppose that
$I_w,I_{w-1},\dots,I_{k+1}$ are already defined. Set $I_k=\emptyset$. For each of the vectors $v_i, i=1,\dots,m$ consider the set
of coordinates $L_{k,i}:=\supp(v_i)\backslash\cup_{i=k}^{w}I_i$, and if the size of $L_{k,i}$ is exactly $k,$ put $I_k\leftarrow I_k\cup L_{k,i}.$

It is easy to observe that each block $I_i$ in the partition $\cI_w$ is a disjoint union of $t=|I_i|/i$ subsets $L_{i,j}$ such that $\abs{L_{i,j}}=j$ for all $j,$ and each $L_{i,j}$ is contained in the support of a different vector $v_{i,j} \in \set{v_l}_{l\in [m]}$ Moreover, the set $\supp(v_{i,j})\setminus L_{i,j}$ is a subset of $\cup_{\ell=k+1}^w I_\ell.$ An important property of the partition $\cI_w$ is as follows. 
\begin{lemma}[{\cite[Lemma 2]{iceland2015coset}}]  Let $A=2/p^w.$ There exists an index $k\in\{1,2,\dots,w\}$ such that
		$$
		|I_k| \geq \max \Big\{ A \sum_{j=k+1}^w |I_j|, \frac{n}{2w A^w} \Big\}.
		$$
	\label{iceland_lemma22} 
\end{lemma}
Now let $k$ be the index whose existence is guaranteed by this lemma. We have $|I_k| \geq \max \big\{ A \sum_{j>k} |I_j|, \frac n{2wA^w} \big\}$ and the set $I_k$ is a disjoint union of $t=|I_k|/k$ sets $L_{k,j}.$ Now consider a coset leader $x$. An easy argument shows that $\supp(x)$ contains no more than $t p^k/2$ subsets $L_{k,j}.$ Indeed, let $S:=\{j|L_{k,j}\subset \supp(x)\}$ and consider a vector $y$ from the same coset given by
\begin{equation}\label{coset_leader_constraint}
	y=x+\sum_{j\in S}v_{k_j}   	
\end{equation}

If $|S|\ge tp^k/2$ then we would obtain $\wt(y)<\wt(x),$ a contradiction. 

The final step is to take a random vector $y$ as in Proposition \ref{lemma:pi} and to estimate the probability that it is a coset leader of the code $\cC^\bot.$ First observe that for $k$ given by Lemma \ref{iceland_lemma22} we have
   $$
   {p^k}t=\frac{p^k}{k}\frac{|I_k|}{n} n\ge \frac{p^w}{w}\frac 1{2w A^w}n=\frac 1{w^2}\Big(\frac{p^w}2\Big)^{w+1}n
      $$  
Now let $Y\sim \text{Binom}(t,p^k),$ then the Chernoff bound gives
   \begin{equation}\label{eq:C}
   \pi_p\le P\Big(Y\le \frac{p^k t}2\Big)\le \exp\Big(-\frac{p^k t}{8}\Big) \le2^{-c(w,p) n},
   \end{equation}
where $c(\cdot,\cdot)$ is defined in Theorem \ref{thm:linear}.
  
Before finishing the proof of Theorem~\ref{thm:linear}, observe that the requirement for $\set{v_i}_{i\in [m]}$ to form a basis of $\cC^\bot$ was not used in the above argument, including the omitted proof of Lemma~\ref{iceland_lemma22} (this proof, due to
\cite{iceland2015coset}, applies here verbatim). The only assumptions used are those listed in \eqref{eq:union}(a)-(b), namely that
there exists a set of low-weight dual codewords whose supports jointly cover all the coordinates.
The last assumption is used in the construction of the partition $\cI_w$ in Lemma~\ref{iceland_lemma22} and in \eqref{coset_leader_constraint}. 

This enables us to use inequalities  \eqref{eq:BB}, and \eqref{eq:C} in \eqref{eq:FT} (taking $p=\tau$ and noting that $\tau<1/2$ for all $\delta\in(0,1/2)$). This substitution proves Lemma \ref{lemma:ldpc}, and the estimate \eqref{eq:ldpc} in Theorem~\ref{thm:linear} follows immediately upon taking $w=r+1.$
	
%	\begin{figure}
%	\center
%	\includegraphics[width=9cm,keepaspectratio]{comparison1.pdf}
%	\caption{Comparison of the asymptotic upper-bound on rate (vs relative minimum distance) in \cref{UB_rate} (blue) and \cref{thm:linear} (red) with the existing bounds in \cite{Gop11,CaMa2015,mceliece1977new} for locality $r=2$.}\label{plot_r2}
%	\end{figure}	
%
%	\begin{figure}
%	\center
%	\includegraphics[width=9cm,keepaspectratio]{improvement.eps}
%	\caption{Difference between the bound in \eqref{eq:asymptoticonverse} (cf. \cite{CaMa2015}) and our bound \cref{UB_rate}.}\label{comparison_table}
%	\end{figure}  

\subsection{Upper Bound on LRC Codes with Disjoint Repair Groups} 
\label{sec:DRG}
Upper bounds for LRC codes with disjoint repair groups were already considered in Sec.~\ref{sect:LP}. Here we consider the asymptotic version of
this problem, noting that the general result of Theorem \ref{thm:linear} in this case admits a significant improvement.
	
Assume that $n=(r+1)m$ and consider a binary linear LRC code $\cC$ of length $n$, minimum distance $\delta n$, and locality $r$ with disjoint repair groups $\cR_j, j=1,\dots,m.$ For every $j$ the repair group $\cR_j$ corresponds to a vector $v_j$ of weight $r+1$ in $\cC^\bot$, and these vectors have pairwise disjoint supports and trivially satisfy the conditions in \eqref{eq:union} (even if some repair groups are of smaller size, we can add redundant coordinates to them to bring their size to $r+1$). 

Recall that $\cB(\tau n)=\cB_\Gamma(0,\tau n)$ is the set of coset
leaders of $\cC^\bot$ of weight at most $\tau n$ and let $x\in\cB(\tau n).$ The vector $x$ satisfies the following constraints:
 	\begin{subequations}\label{constraint1}
 	\begin{equation}
	   \wt(x)\le \min_{T\subset[m]} \wt\Big( x+\sum_{i\in T}v_i\Big), \label{constraint1_minwt}
	   \end{equation}
	   \begin{equation}
 		\wt(x) \leq \tau n,  \label{constraint1_maxwt}
		\end{equation}
		\begin{align}\label{constraint1_uniqueness}
		\begin{split}
		|\cB(\tau n)\cap (x+\spn(\{v_i, i=1,\dots,m\}))|\le 1\\ \text{for all }x \in\{0,1\}^n.
		\end{split}
		 \end{align}
 	\end{subequations}
  Eqns.~\eqref{constraint1_minwt}-\eqref{constraint1_uniqueness} can be used to derive an upper bound on $R(\cC)$ which is given in 
  the following theorem.

	\begin{figure*}
	\center
	\includegraphics[width=16cm,keepaspectratio]{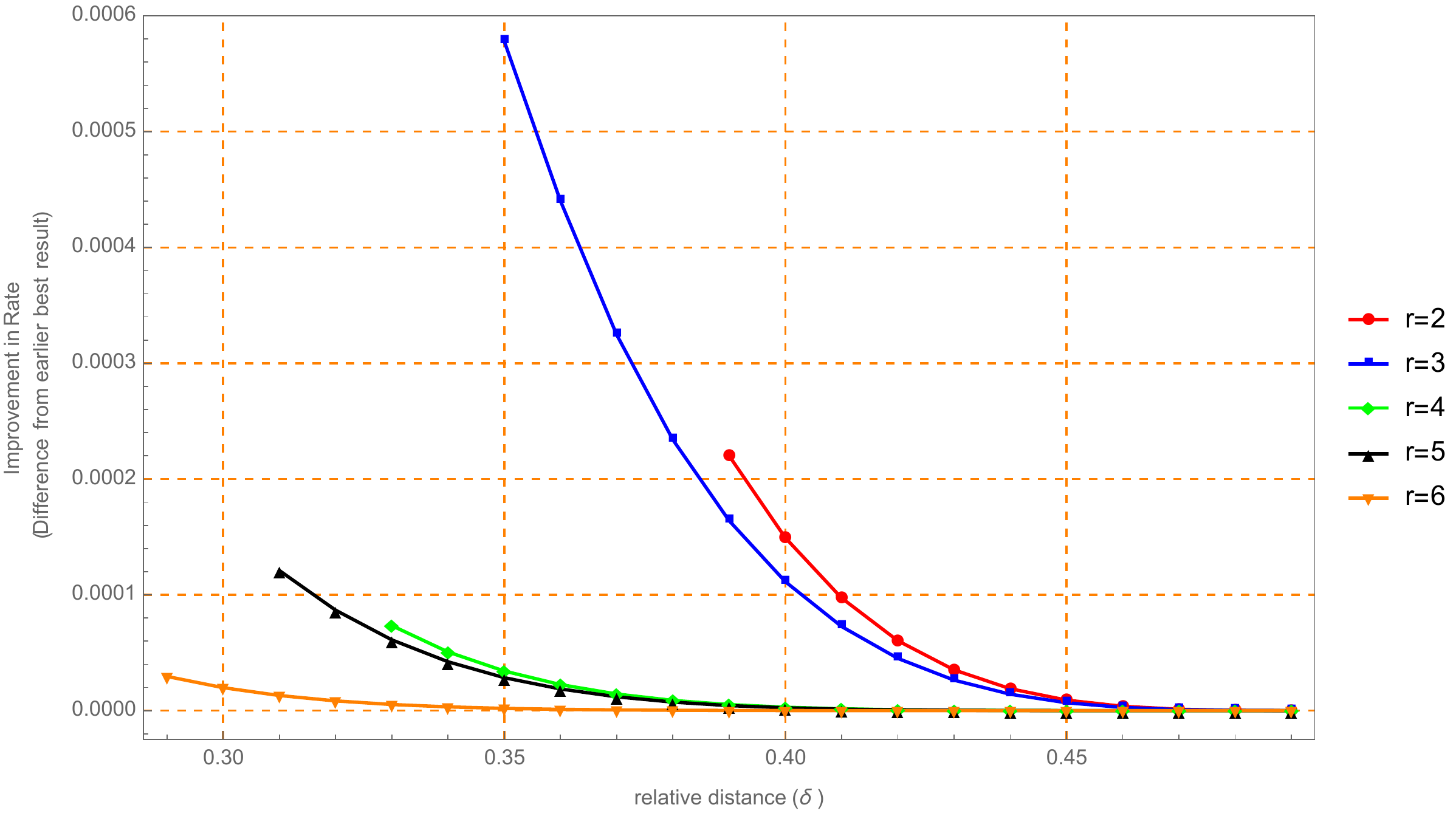}
	\caption{The difference between the bound on rate in  \eqref{UB_rate} and the best previously known result \eqref{eq:upper} (see \cite{CaMa2015}). It can be noted that, as $r$ increases, the improvement over \eqref{eq:upper} diminishes.}
		\label{fig:compar}
	\end{figure*}    

\begin{theorem}[Linear LRC codes with disjoint repair groups]\label{thm:disjoint}
Let $R^{\text{\rm  (lin,dis)}}(r,\delta)$ be the largest possible asymptotic rate of linear LRC codes with disjoint repair groups.
Let $t \defeq \floor*{\frac{r+1}{2}}$. We have
       \begin{equation}\label{UB_rate}
		R^{\text{\rm  (lin,dis)}}(r,\delta) \leq  \tau \Log\br{x} + \frac{\log\br{\sum_{i=0}^{t}\beta_i(x)}}{r+1} \Big\vert_{x=\mu}
  	\end{equation}
where
     	\begin{equation}\label{def_beta_i}  		
  	\beta_i(x) \defeq \begin{cases}
  		\displaystyle{\frac{\binom{r+1}i}{x^i}} & \text{if }i<t \\[.1in]
  		\displaystyle{\frac{\binom{r+1}i}{2^{r\mmod\! 2}}\, x^i} & \text{if }i=t,
  	\end{cases}
  	\end{equation}
where	$\mu =  1$
	if $\sum_{i=0}^t  i \frac{\beta_i(x)}{\sum_\ell \beta_\ell(x)}\Big\vert_{x=1} \leq (r+1)\tau$ and
	\begin{align}\label{def_mu} 
	\begin{split}
\mu =   {\text{\rm Root}}^+\Big((r+1) \tau \sum_{i=0}^t \binom{r+1}i x^{t-i}  \\
- \sum_{i=0}^t \binom{r+1}i i\; x^{t-i}\Big)
\end{split}
 \end{align}
  otherwise. 
Here ${\text{\rm Root}}^+(f(x))$ denotes the unique positive zero of the polynomial $f(x).$
\end{theorem}
\begin{IEEEproof}	We again rely on inequality \eqref{eq:FT}. To bound above the $B(\tau n)$ let us the constraints in \eqref{constraint1}.
In particular, from  \eqref{constraint1_minwt} and \eqref{constraint1_maxwt} we obtain
	\begin{align}\label{UB_CLset}
	\begin{split}
		\cB(\tau n) \subseteq \Big\{x\in &\{0,1\}^n \Big| \wt(x)\le \tau n;\\
		& |\supp(x)\cap \cR_i|\le t \text{ for all $i$}\Big\}.
		\end{split} 
	\end{align}
	The number of vectors $x$ that satisfy \eqref{constraint1_minwt}-\eqref{constraint1_maxwt} is therefore given by the coefficient of
	$x^k$ in the following expression:
  \begin{equation}\label{generating_fn}
		\sum_{j\leq \tau n} \text{Coeff\,}_{x^j} \Big({1+\sum_{i=0}^{t}\binom{r+1}i x^i}\Big)^{m}.
	\end{equation}
Now let us in addition use \eqref{constraint1_uniqueness}.
Since $\wt(x) \leq \wt(x+v_i)$ for all $i$ and they are in the same coset, it suffices to count only one of them on the right-hand side of 
\eqref{UB_CLset}. Therefore if $r+1$ is even, we obtain
   $$
   B(\tau n)\le \sum_{j=0}^\tau n\, \text{Coeff\,}_{x^j} (g(x))
   $$
   where
   $$
   g(x)=\Big({1+\sum_{i=0}^{t}\binom{r+1}i x^i}+ {\frac{\binom{r+1}i}{2^{r\mmod\! 2}}\, x^i}\Big)^{m}
   $$
One can see that asymptotically for $n\to\infty$ the dominating term in this expression is given by the largest coefficient, i.e., 
  \begin{align*}
  \frac1n\log{\displaystyle\sum_{0\le j\leq \tau n}\text{Coeff\,}_{x^j}(  g(x))}
           \sim \frac 1n \max_{0\le j\leq \tau n} \log \text{Coeff\,}_{x^j} (g(x)).
   \end{align*}
	Thus we have
	  $$
	  R^{\text{\rm  (lin,dis)}}(r,\delta) \le \max_{0\le j\leq \tau n}{\log \text{Coeff\,}_{x^j}(  g(x))}
	  $$
which translates into the following convex maximization problem:
  $$
\max \quad  \frac1 {r+1}H(P) + \sum_{i=0}^{t-1} \alpha_i \log{\binom{r+1}i}+ \alpha_t \frac{\binom{r+1}i}{2^{r\mmod\! 2}} 
 $$
where the distribution $P=(r+1)(\alpha_0,\dots,\alpha_t)$ satisfies the constraints
  $$
    \sum_{i=0}^t\alpha_i = \frac{1}{r+1};\;\; \sum_{i=0}^t i \alpha_i \leq \tau  
  $$   
 The maximum is found by differentiation (setting up a Lagrange function) and we obtain 
   $$
   \alpha_i = \frac{\beta_i(x)}{(r+1)\sum_{i=1}^t \beta_i(x)}\Big\vert_{x=\mu}, \quad i=0,1,\dots,t
   $$
for  $\beta_i(x)$ and $\mu$ as defined in \eqref{def_beta_i},\eqref{def_mu}. Substituting these values of $\alpha_i,$ we find
the bound in the form given in \eqref{UB_rate}.
 \end{IEEEproof}

A numerical evaluation of the improvements in the asymptotic rate of this results over the previous existing results is shown in 
Fig.~\ref{fig:compar}.
%\cref{table1,table2}
The improvement can be seen for larger values of the relative distance. For instance, for $r=2$ the improvement is obtained for $\delta \geq 0.38,$  and this range increases for larger values of $r$. % (the improvements are confirmed numerically and are rather small).

%\begin{table}[ht] 
% \centering 
%    \caption{Improvements in the asymptotic rate for Thm.~ \ref{thm:linear} over the existing bounds in \eqref{eq:upper} from \cite{CaMa2015} for r = 2.} 
%    \begin{tabular}{|c|c|c|c|c|c|c|c|c|c|c|} \hline 
%      $\delta$ & 0.04   &  0.09  &   0.14   &   0.19   &  0.24 &  0.29 & 0.34 & 0.39  & 0.44 & 0.49 \\\hline 
%      Diff  & -0.0586 & -0.118 & -0.113 & -0.085 & -0.0532 & -0.0251 & -0.00587 & 0.00022 & & 5\\  \hline
%    \end{tabular} 
%\end{table} 

%\begin{figure}
%    \centering
%    \begin{subfigure}[b]{0.3\textwidth}
%        \includegraphics[width=\textwidth]{table1.pdf}
%        \caption{Improvements in the asymptotic rate for \cref{thm:linear} over the existing bounds in \cite{CaMa2015} for $r=2$}
%        \label{table1}
%    \end{subfigure}
%    ~ \qquad\qquad\qquad%add desired spacing between images, e. g. ~, \quad, \qquad, \hfill etc. 
%      %(or a blank line to force the subfigure onto a new line)
%    \begin{subfigure}[b]{0.3\textwidth}
%        \includegraphics[width=\textwidth]{table2.pdf}
%        \caption{Improvements in the asymptotic rate for \cref{thm:linear} over the existing bounds in \cite{CaMa2015} for $r=2$}
%        \label{table2}
%    \end{subfigure}
%  \end{figure}  

\vspace*{.1in}{\sc Acknoledgment.} We are grateful to the reviewers for insightful remarks that helped us to improve the presentation of the paper. In particular, Lemma \ref{lemma:linear} was suggested by a reviewer.

\bibliographystyle{IEEEtran}
\bibliography{CombBounds}

\begin{thebibliography}{10}
\providecommand{\url}[1]{#1}
\csname url@samestyle\endcsname
\providecommand{\newblock}{\relax}
\providecommand{\bibinfo}[2]{#2}
\providecommand{\BIBentrySTDinterwordspacing}{\spaceskip=0pt\relax}
\providecommand{\BIBentryALTinterwordstretchfactor}{4}
\providecommand{\BIBentryALTinterwordspacing}{\spaceskip=\fontdimen2\font plus
\BIBentryALTinterwordstretchfactor\fontdimen3\font minus
  \fontdimen4\font\relax}
\providecommand{\BIBforeignlanguage}[2]{{%
\expandafter\ifx\csname l@#1\endcsname\relax
\typeout{** WARNING: IEEEtran.bst: No hyphenation pattern has been}%
\typeout{** loaded for the language `#1'. Using the pattern for}%
\typeout{** the default language instead.}%
\else
\language=\csname l@#1\endcsname
\fi
#2}}
\providecommand{\BIBdecl}{\relax}
\BIBdecl

\bibitem{Gop11}
P.~Gopalan, C.~Huang, H.~Simitci, and S.~Yekhanin, ``On the locality of
  codeword symbols,'' \emph{IEEE Trans. Inform. Theory}, vol.~58, no.~11, pp.
  6925--6934, 2011.

\bibitem{PKLK2012}
N.~Prakash, G.~M. Kamath, V.~Lalitha, and P.~V. Kumar, ``Optimal linear codes
  with a local-error-correction property,'' in \emph{Proc. 2012 IEEE Internat.
  Sympos. Inform. Theory}, 2012, pp. 2776--2780.

\bibitem{kamath2013}
G.~M. Kamath, N.~Prakash, V.~Lalitha, and P.~V. Kumar, ``Codes with local
  regeneration,'' in \emph{Proc. IEEE Int. Symp. Inform. Theory, Istanbul,
  Turkey, Jul. 2013}, pp. 1606--1610.

\bibitem{Lluis2013}
L.~Pamies-Juarez, H.~D.~L. Hollmann, and F.~E. Oggier, ``Locally repairable
  codes with multiple repair alternatives,'' in \emph{Proc. 2013 IEEE Int.
  Sympos. Inform. Theory}, pp. 892--896.

\bibitem{wang2014a}
A.~Wang and Z.~Zhang, ``Repair locality with multiple erasure tolerance,''
  \emph{IEEE Trans. Inform. Theory}, vol.~60, no.~11, pp. 6979--6987, Nov 2014.

\bibitem{Prakash14}
N.~Prakash, V.~Lalitha, and P.~Kumar, ``Codes with locality for two erasures,''
  in \emph{Proc. 2014 IEEE Int. Sympos. Inform. Theory, Honolulu, HI}, pp.
  1962--1966.

\bibitem{RawatEurasip15}
A.~S. Rawat, A.~Mazumdar, and S.~Vishwanath, ``Cooperative local repair in
  distributed storage,'' \emph{{EURASIP} Journal on Advances in Signal
  Processing}, 2015, 17pp.

\bibitem{mazumdar2015storage}
A.~Mazumdar, ``Storage capacity of repairable networks,'' \emph{IEEE
  Transactions on Information Theory}, vol.~61, no.~11, pp. 5810--5821, 2015.

\bibitem{sil13}
N.~Silberstein, A.~S. Rawat, O.~Koyluoglu, and S.~Vishwanath, ``Optimal locally
  repairable codes via rank-metric codes,'' in \emph{Proc. IEEE Int. Sympos.
  Inform. Theory, Boston, MA}, 2013, pp. 1819--1823.

\bibitem{huang2007pyramid}
C.~Huang, M.~Chen, and J.~Li, ``Pyramid codes: Flexible schemes to trade space
  for access efficiency in reliable data storage systems,'' in \emph{Sixth IEEE
  International Symposium on Network Computing and Applications}, 2007, pp.
  79--86.

\bibitem{TPDMatroids}
I.~Tamo, D.~S. Papailiopoulos, and A.~G. Dimakis, ``Optimal locally repairable
  codes and connections to matroid theory,'' in \emph{Proc. 2013 IEEE Int.
  Sympos. Inform. Theory}, 2013, pp. 1814--1818.

\bibitem{Tamo13}
I.~Tamo and A.~Barg, ``A family of optimal locally recoverable codes,''
  \emph{IEEE Trans. Inform. Theory}, vol.~60, no.~8, pp. 4661--4676, 2014.

\bibitem{BTV17}
A.~Barg, I.~Tamo, and S.~Vl{\u{a}}du{\c{t}}, ``Locally recoverable codes on
  algebraic curves,'' \emph{IEEE Trans. Inform. Theory}, vol.~63, no.~8, pp.
  4928--4939, 2017.

\bibitem{barg15}
A.~Barg and I.~Tamo, ``On codes with the locality property,'' \emph{IEEE
  Information Theory Society Newsletter}, vol.~65, no.~4, pp. 7--15, 2015.

\bibitem{kamath14}
G.~M. Kamath, N.~Prakash, V.~Lalitha, and P.~V. Kumar, ``Codes with local
  regeneration and erasure correction,'' \emph{IEEE Trans. Inform. Theory},
  vol.~60, no.~8, pp. 4637--4660, 2014.

\bibitem{raw14}
A.~Rawat, D.~Papailiopoulos, A.~Dimakis, and S.~Vishwanath, ``Locality and
  availability in distributed storage,'' \emph{IEEE Trans. Inform. Theory},
  vol.~62, no.~8, pp. 4481--4493, 2016.

\bibitem{TBF16}
I.~Tamo, A.~Barg, and A.~Frolov, ``Bounds on the parameters of locally
  recoverable codes,'' \emph{IEEE Trans. Inform. Theory}, vol.~62, no.~6, pp.
  3070--3083, 2016.

\bibitem{CaMa2015}
V.~Cadambe and A.~Mazumdar, ``Bounds on the size of locally recoverable
  codes,'' \emph{IEEE Trans. Inform. Theory}, vol.~61, no.~11, pp. 5787--5794,
  2015.

\bibitem{del73}
P.~Delsarte, ``An algebraic approach to the association schemes of coding
  theory,'' \emph{Philips Research Repts Suppl.}, vol.~10, pp. 1--97, 1973.

\bibitem{mce77}
R.~McEliece, E.~Rodemich, H.~Rumsey, and L.~Welch, ``New upper bounds on the
  rate of a code via the {D}elsarte-{M}ac{W}illiams inequalities,''
  \emph{Information Theory, IEEE Transactions on}, vol.~23, no.~2, pp.
  157--166, 1977.

\bibitem{ChanLP13}
T.~H. Chan, M.~A. Tebbi, and C.~W. Sung, ``Linear programming bounds for
  storage codes,'' in \emph{9th International Conference on Information,
  Communications and Signal Processing (ICICS)}, 2013, pp. 1--5.

\bibitem{TebbiChanSung2014}
M.~A. Tebbi, T.~H. Chan, and C.~W. Sung, ``Linear programming bounds for robust
  locally repairable storage codes,'' in \emph{2014 IEEE Information Theory
  Workshop (ITW 2014)}, Nov 2014, pp. 50--54.

\bibitem{CyclicLRC16}
I.~Tamo, A.~Barg, S.~Goparaju, and R.~Calderbank, ``Cyclic {LRC} codes, binary
  {LRC} codes, and upper bounds on the distance of cyclic codes,''
  \emph{International Journal of Information and Coding Theory}, vol.~3, no.~4,
  pp. 345--364, 2016.

\bibitem{iceland2015coset}
E.~Iceland and A.~Samorodnitsky, ``On coset leader graphs of {LDPC} codes,''
  \emph{IEEE Trans. Inform. Theory}, vol.~61, no.~8, pp. 4158--4163, 2015.

\bibitem{HTB16}
S.~Hu, I.~Tamo, and A.~Barg, ``Combinatorial and {LP} bounds for {LRC} codes,''
  in \emph{Proc IEEE Int. Sympos. Inform. Theory}, Barcelona, Spain, 2016, pp.
  1008--1012.

\bibitem{AM16}
A.~Agarwal and A.~Mazumdar, ``Bounds on the rate of linear locally repairable
  codes over small alphabets,'' July 2016, arXiv:1607.08547.

\bibitem{bro89}
A.~E. Brouwer, A.~M. Cohen, and A.~Neumaier, \emph{Distance-regular
  graphs}.\hskip 1em plus 0.5em minus 0.4em\relax Berlin e. a.:
  Springer-Verlag, 1989.

\bibitem{MaSl1977}
F.~J. MacWilliams and N.~J.~A. Sloane, \emph{The theory of error-correcting
  codes}.\hskip 1em plus 0.5em minus 0.4em\relax North-Holland Publishing Co.,
  Amsterdam e.a., 1977.

\bibitem{BHPS1998}
A.~E. Brouwer, H.~O. H{\"a}m{\"a}l{\"a}inen, P.~R.~J. {\"O}sterg{\aa}rd, and
  N.~J.~A. Sloane, ``Bounds on mixed binary/ternary codes,'' \emph{IEEE Trans.
  Inform. Theory}, vol.~44, no.~1, pp. 140--161, 1998.

\bibitem{Ma1999}
W.~J. Martin, ``Designs in product association schemes,'' \emph{Des. Codes
  Cryptogr.}, vol.~16, no.~3, pp. 271--289, 1999.

\bibitem{sage}
{S}age {M}athematics {S}oftware ({V}ersion 6.9), 2015, http://www.sagemath.org.

\bibitem{Cal85}
A.~Calderbank and J.-M. Goethals, ``On a pair of dual subschemes of the hamming
  scheme {$H_n(q)$},'' \emph{Europ. J. Combinatorics}, vol.~6, pp. 133--147,
  1985.

\bibitem{del91}
C.~Delorme and P.~Sol{\'e}, ``Diameter, covering index, covering radius and
  eigenvalues,'' \emph{European J. Comb.}, vol.~12, no.~2, pp. 95--108, 1991.

\bibitem{friedman2005generalized}
J.~Friedman and J.-P. Tillich, ``Generalized {A}lon--{B}oppana theorems and
  error-correcting codes,'' \emph{SIAM Journal on Discrete Mathematics},
  vol.~19, no.~3, pp. 700--718, 2005.

\end{thebibliography}

\end{document}